\documentclass[aps,superscriptaddress,showpacs,nofootinbib, 10pt]{revtex4-1}
\usepackage{mathrsfs}
\usepackage{bigints}
\usepackage{latexsym,bm}
\usepackage{graphicx}
\usepackage{indentfirst}
\usepackage{slashed}
\usepackage{amssymb}
\usepackage{amsmath}
\usepackage{bbm}
\usepackage{color}
\usepackage[dvipsnames]{xcolor}
\usepackage{epsfig}
\usepackage[titletoc]{appendix}
\usepackage{multirow}%
\usepackage{rotating}
\usepackage{epstopdf}
\usepackage{extarrows}
\usepackage{tabularx}
\usepackage{soul} 
\usepackage{xcolor}
\usepackage{lipsum}
\usepackage[colorlinks=true,allcolors=BlueViolet,hyperfootnotes=false]{hyperref}

\usepackage[utf8]{inputenc}

\newcommand{\be}{\begin{equation}} \newcommand{\ee}{\end{equation}}
\newcommand{\ba}{\begin{array}{c}} \newcommand{\ea}{\end{array}}
\newcommand{\bea}{\begin{eqnarray}} \newcommand{\eea}{\end{eqnarray}}

\newcommand\tstrut{\rule{0pt}{2.9ex}}       

\begin{document}
\title{\Large Hadron and lepton tensors in semileptonic decays including new physics.}

\author{Neus Penalva}
\affiliation{Departamento de F\'\i sica Fundamental,\\ Universidad de Salamanca, E-37008 Salamanca, Spain}

\author{Eliecer Hern\'andez}
\affiliation{Departamento de F\'\i sica Fundamental 
  e IUFFyM,\\ Universidad de Salamanca, E-37008 Salamanca, Spain}

\author{Juan~Nieves}
\affiliation{Instituto de F\'{\i}sica Corpuscular (centro mixto CSIC-UV), Institutos de Investigaci\'on de Paterna,
Apartado 22085, 46071, Valencia, Spain}

\date{\today}
\begin{abstract}
 We extend our general framework for semileptonic decay,  originally
  introduced in
N. Penalva et al. [Phys. Rev. D100, 113007 (2019)], with the 
addition of  new physics (NP) tensor terms. In this way, 
all the NP effective Hamiltonians
that are 
 considered in lepton flavour universality violation (LFUV) studies
 have now been  included. Those  are
 left and right vector and scalar NP hamiltonians and the NP tensor one. Besides, 
we now also give general expressions that allow for complex Wilson coefficients. 
The scheme developed is totally general and it can be
 applied to any charged current semileptonic decay, involving any quark 
 flavors or initial and final hadron states. We show that all the hadronic input, including NP effects,  can be parametrized in 
 terms of  16  Lorentz scalar structure functions, 
 constructed out of  the NP complex Wilson coefficients and the  genuine hadronic responses, with the latter determined 
 by the matrix elements of the involved hadron operators.  In the second part  of this work,  we use this formalism to 
obtain the complete NP effects in the $\Lambda_b\to \Lambda_c \tau\bar\nu_\tau$ semileptonic decay, where LFUV, if finally confirmed,
is also expected to be seen. We stress the relevance of the 
center of mass (CM) $d^2\Gamma/(d\omega d\cos\theta_\ell)$ and laboratory (LAB) $d^2\Gamma/
(d\omega dE_\ell)$ differential decay widths, with  $\omega$   the product 
of the hadron four-velocities,  $\theta_\ell$  the angle made 
by the three-momenta of 
the charged lepton and the final hadron in the $W^-$ CM frame and 
$E_\ell$  the charged lepton energy in the decaying hadron
rest frame.  While models with very different strengths 
in the NP terms give the same differential $d\Gamma/d\omega$ and total decay 
widths for this decay, they predict very different numerical results for some of the $\cos\theta_\ell$ and $E_\ell$
coefficient-functions that determine the above two distributions. Thus,  the combined analysis of the CM $d^2\Gamma/(d\omega d\cos\theta_\ell)$ and 
LAB  $d^2\Gamma/ (d\omega dE_\ell)$ differential decay widths  will  help clarifying what kind
of NP is a better candidate  in order to explain LFUV.
\end{abstract}

\maketitle
\section{Introduction}
\vspace{1cm}
Present discrepancies, between  experimental data and  the Standard Model (SM) 
results, seen 
 in semileptonic $B-$meson decays, point at the possible existence of new physics (NP). 
 Since there is no evidence of similar discrepancies in transitions involving 
 the two first quark and lepton families, it is generally assumed that the
possible  NP contributions  only affect  the third quark and lepton generations,
being thus  responsible for  lepton flavor universality violation (LFUV). These violations
 have been  seen  in the values of the $\Gamma(B\to D)$ and $\Gamma(B\to D^*)$ semileptonic decays
 widths when compared to SM predictions 
 (see for instance Ref.~\cite{Bifani:2018zmi}). 
  These discrepancies are presented in form of ratios
${\cal R}_{D^{(*)}}=\frac{\Gamma(B\to D^{(*)}\tau\bar\nu_\tau)}{\Gamma(B\to
D^{(*)}\ell\bar\nu_\ell)}$, with  $\ell=e,\mu$, for which part
 of the hadronic uncertainties should cancel.
 While very recent preliminary measurements by the Belle
Collaboration~\cite{Abdesselam:2019dgh,Belle:2019rba} reduce  the tension with the SM 
 to the level of $0.8\sigma$, a combined analysis of 
BaBar~\cite{Lees:2012xj,Lees:2013uzd}, 
Belle~\cite{Huschle:2015rga,Sato:2016svk,Hirose:2016wfn} 
and LHCb~\cite{Aaij:2015yra,Aaij:2017uff} data and SM 
predictions~\cite{Aoki:2016frl,Aaij:2015yra, Bigi:2016mdz, Jaiswal:2017rve}
by the Heavy Flavour Averaging Group (HFLAV)~\cite{Amhis:2016xyh} shows a tension
with the SM at the level of $4.4\sigma$. This tension reduces to $3.1\sigma$ 
if the latest Belle results are taken into 
account~\cite{Amhis:2019ckw}.
From the theoretical point of view, phenomenological  approaches follow an
effective field theory model-independent analysis that includes  
different $b\to c \ell \nu_\ell$ 
charged current (CC) effective operators. They are assumed to be generated 
by physics beyond the SM, that   would enter at a much higher energy scale, and their strengths
  are given by  unknown Wilson coefficients that should be  fitted to data.
 Those analyses include the  pioneering work of Ref.~\cite{Fajfer:2012vx} or the more recent one in
Ref.~\cite{Murgui:2019czp} from which we shall take the values for the Wilson coefficients 
that we are going to use in the numerical part of this work.

The anomaly seen in $B$ decays could be corroborated  in other processes 
governed by the same $b\to c$ transition like the
$\Lambda_b\to\Lambda_c\ell\bar\nu_\ell$ decays. The LHCb Collaboration~\cite{Aaij:2017svr}
 has very recently measured the shape of the $d\Gamma(\Lambda_b\to\Lambda_c
 \mu^-\bar\nu_\mu)/d\omega$ decay width, and it is expected~\cite{Cerri:2018ypt}
that the  precision in the ${\cal
R}_{\Lambda_c}=\frac{\Gamma(\Lambda_b\to\Lambda_c\tau\bar\nu_\tau)}{
\Gamma(\Lambda_b\to\Lambda_c\mu\bar\nu_\mu)}$ ratio might
reach that obtained for ${\cal R}_D$ and ${\cal R}_{D^*}$.
Heavy quark
spin symmetry (HQSS) strongly constraints
the form factors relevant for this
transition, with no subleading Isgur-Wise (IW) 
function occurring at order ${\cal O}(\Lambda_{QCD}/m_{b,c})$, and only two 
subleading ones entering at next order
 \cite{Neubert:1993mb,Bernlochner:2018bfn,Bernlochner:2018kxh}.  
The ${\cal R}_{\Lambda_c}$ ratio has been accurately predicted within the SM
in Ref.~\cite{Bernlochner:2018kxh} with the use of leading and subleading
 HQSS IW functions that were simultaneously fitted to LQCD 
results and LHCb data.  Precise results for the vector and axial form factors 
  were obtained in  Ref.~\cite{Detmold:2015aaa} using Lattice
QCD (LQCD) with 2+1 flavors of dynamical domain-wall fermions. 
The additional form factors needed to include NP tensor terms have been obtained 
within the same LQCD scheme in Ref.~\cite{Datta:2017aue}.  The  also needed
 scalar  and pseudoscalar  form factors  can be directly related to vector and axial ones
(see  Eqs.~(2.12) and (2.13) of Ref.~\cite{Datta:2017aue}).
 With a lot of theoretical effort involved
\cite{Li:2016pdv,Blanke:2018yud,Azizi:2018axf,Blanke:2019qrx,Boer:2019zmp,
Shivashankara:2015cta,
Datta:2017aue, Bernlochner:2018bfn,Ray:2018hrx, 
 DiSalvo:2018ngq, Murgui:2019czp,Ferrillo:2019owd,Penalva:2019rgt} in checking
  the effects of NP scenarios and with 
expectations of experimental data in the near future,
 this reaction could also play an important role in the study of $b\to c$ LFUV 
studies.

In this work we introduce  a general framework to study any baryon/meson  
semileptonic decay for unpolarized hadrons including NP contributions, 
although we will refer explicitly only to those decays induced by a $b\to c$ 
transition. We consider a general scheme, based in the co called Standard Model 
Effective Field
Theory (SMEFT) scheme~\cite{Buchmuller:1982ye,Grzadkowski:2010es}, to 
analyze any decay driven by a $q\to q' \ell \bar\nu_\ell$ quark level 
CC  process involving massless left-handed neutrinos. 
 We allow for CP--violating scalar, pseudo-scalar and tensor NP terms, 
 as well as corrections to the SM vector and axial contributions. 
 All the hadronic input, including NP effects,  can be parametrized in 
 terms of  16  Lorentz scalar structure functions $\widetilde W's$ (SFs), 
 constructed out of  NP complex Wilson coefficients 
 ($C's$) and the  genuine hadronic responses ($W's$), which are determined 
 by the matrix elements of the involved hadron operators. The $W$ 
 SFs\footnote{Symbolically, $\widetilde W = C W $.}  depend on the masses 
 of the initial and final particles and on the invariant mass ($q^2$) of the 
 outgoing  $\ell \nu_\ell$ pair, and they can be expressed in terms of the form-factors 
 used to parametrize the transition matrix elements.  
 
 In the case of the SM they reduce to  just five real $\widetilde W $ 
SFs and, provided that massless ($e$ or $\mu$) and
  $\tau-$ mode decays are 
simultaneously analyzed, all five  $\widetilde W $ 
SFs can be determined either from the unpolarized 
$d^2\Gamma/(d\omega d\cos\theta_\ell)$
decay width, where $\omega$ is the product of the two hadron four velocities 
and $\theta_\ell$ the angle made by the final hadron and  charged lepton 
three-momenta in 
the center of mass of the two final leptons (CM), or from the unpolarized 
$d^2\Gamma/(d\omega dE_\ell)$ decay width, where $E_\ell$ is the charged lepton energy 
 measured in the laboratory system (LAB).

The unpolarized CM $d^2\Gamma/(d\omega d\cos\theta_\ell)$ and 
LAB $d^2\Gamma/(d\omega dE_\ell$ decay widths get contributions 
from both positive and negative charged lepton helicities, contributions
that  have also  been explicitly evaluated  in this work. Assuming NP, these new observables are  sensitive to new 
combinations of the $\widetilde W$ SFs, and thus serve to further restrict  the relevance of operators beyond the SM. 
There are a total of five new independent linear combination of the $\widetilde W$ SFs needed to describe
the case with a polarized final charged lepton. To determine them, the 
LAB and CM charged lepton helicity distributions have to be used  simultaneously, since in this case they
 provide complementary  information.

As mentioned, we have considered five NP Wilson coefficients. In general 
they are complex, although one of them can always be taken to be real. 
Therefore, nine free parameters should be determined from data. Even assuming 
that the form factors are known, and therefore the genuinely hadronic part ($W$) of
 the $\widetilde W$ SFs, all NP parameters  are difficult to be determined
  from a  unique type of decay, since  the  experimental measurement of the required 
  polarized decay is an extremely  difficult task.  It is therefore more convenient to analyze data from various 
types of semileptonic decays simultaneously (f.e. $\bar B\to D, \bar B\to D^*$,
 $\Lambda_b\to \Lambda_c, \bar B_c\to \eta_c, \bar B_c\to J/\Psi$...), 
 considering both the $e/\mu$ and $\tau$ modes. The scheme presented in 
 this work constitutes a powerful tool to achieve this objective.

Besides, within the present framework, it is not difficult to 
consider NP effects induced by light right-handed neutrinos, without including new 
SFs, since we give general expressions for all the hadron tensors. The recent analysis of Ref.~\cite{Mandal:2020htr}, using 
the $b\to c\tau\bar\nu_\tau$ anomalies data in the meson sector, does not rule out NP operators which can arise due to the presence of right-handed neutrinos in the theory, and therefore points to one natural continuation of this work.

Moreover, we stress that all expressions are general and they can be
 applied to any charged current semileptonic decay, involving any quark 
 flavors or initial and final hadron states. Thus for instance, the scheme 
 presented here can also be used to search for NP signatures in nuclear 
 beta decays, from which $|V_{ud}|$ is also determined.

This work is an update of the formalism in Ref.~\cite{Penalva:2019rgt},
 where NP tensor terms were not considered and the CP--conserving limit
  was adopted, assuming that all NP Wilson coefficients were real. It is organized as follows: In Sec.~\ref{sec:forml} we give all the general
 formulae, including the expressions for the effective NP hamiltonians
and the  CM $d^2\Gamma/(d\omega d\cos\theta_\ell)$ and 
LAB $d^2\Gamma/(d\omega dE_\ell)$ differential decay widths both for unpolarized as well
as polarized final charged leptons. Next, in Sec.~\ref{sec:hme}, and to illustrate the general procedure,  
we explain in detail how Lorentz, parity and time-reversal transformations 
constraint  the number of SFs needed to describe the hadronic tensor originating from vector and 
axial current interactions terms. Details for the leptonic tensors and the 
rest of the hadronic tensors are compiled, respectively,  in Appendixes~\ref{sec:appL} and~\ref{sec:appH}. 
In Sec.~\ref{sec:LbtoLc} we introduce the form factors needed for the $\Lambda_b\to\Lambda_c\tau\bar\nu_\tau$ decay 
and apply to this transition the general formalism derived in Sec.~\ref{sec:forml}. We show numerical results using the LQCD form-factors of 
Refs.~\cite{Detmold:2015aaa, Datta:2017aue} and the best-fit Wilson coefficients obtained in \cite{Murgui:2019czp}.  
Conclusions 
are presented in Sec.~\ref{sec:conclusions}. Other relevant information is compiled in Appendixes~\ref{sec:scalaP} (CM and LAB
kinematics), \ref{app:coeff} (expressions for the $\cos\theta_\ell$ and $E_\ell$
coefficient-functions appearing in the CM and LAB distributions in terms of the $\widetilde W$ SFs) 
and \ref{app:ff} (form factors for the  $\Lambda_b\to\Lambda_c\tau\bar\nu_\tau$ transition and general expressions of the
$\widetilde W$ SFs for this decay in terms of the form factors).
\section{Formalism}
\label{sec:forml}
\subsection{Effective Hamiltonian}
\label{sec:eh}

In the context of the SMEFT, we consider  the  effective 
Hamiltonian~\cite{Murgui:2019czp}
\bea
H_{\rm eff}&=&\frac{4G_F|V_{cb}|^2}{\sqrt2}[(1+C_{V_L}){\cal O}_{V_L}+
C_{V_R}{\cal O}_{V_R}+C_{S_L}{\cal O}_{S_L}+C_{S_R}{\cal O}_{S_R}
+C_{T}{\cal O}_{T}]+h.c.,
\label{eq:hnp}
\eea
with fermionic operators given by ($\psi_{L,R}=\frac{1 \mp \gamma_5}{2}\psi$)
\be
{\cal O}_{V_{L,R}} = (\bar c \gamma^\mu b_{L,R}) 
(\bar \ell_L \gamma_\mu \nu_{\ell L}), \quad {\cal O}_{S_{L,R}} = 
(\bar c\,  b_{L,R}) (\bar \ell_R  \nu_{\ell L}), \quad {\cal O}_{T} = 
(\bar c\, \sigma^{\mu\nu} b_{L}) (\bar \ell_R \sigma_{\mu\nu} \nu_{\ell L}).
\label{eq:hnp2}
\ee
The Wilson coefficients $C_i$, complex in general, 
parametrize possible deviations from the SM, i.e. $C_i^{\rm SM}$=0, and 
could be in general, lepton
and flavour dependent, though in Ref.~\cite{Murgui:2019czp} they are assumed to be 
present only in the third generation
of leptons.

\subsection{Decay rate including NP terms}
The  semileptonic differential decay rate of a bottomed hadron
($H_b$) of mass $M$ into a charmed one ($H_c$) of mass $M'$ and 
$\ell \bar\nu_\ell$, measured in its rest frame, 
and after averaging (summing) over the initial (final) hadron polarizations, 
reads\footnote{We emphasize once again that all equations are valid for any   $q \to q '\ell \bar \nu_\ell $ CC decay, although 
we only give explicit expressions for $b \to c $ reactions.}~\cite{Tanabashi:2018oca},
\begin{equation}
  \frac{d^2\Gamma}{d\omega ds_{13}} = \frac{G^2_F|V_{cb}|^2
    M'^2}{(2\pi)^3 M}\, \overline\sum\  |{\cal M}|^2, \label{eq:defsec}
  \end{equation}
where $G_F=1.166\times 10^{-5}$~GeV$^{-2}$  is the Fermi coupling
constant and  ${\cal M} (k,k',p,q,\,{\rm spins} )$ is the transition 
matrix element, with $p$, $k'$, $k=q-k'$ and $p'=p-q$, the decaying 
$H_b$ particle,  outgoing charged lepton,  neutrino and final hadron 
four-momenta, respectively. In addition, $\omega$ is  the product of the two hadron 
four velocities $\omega=(p\cdot p')/(MM')$, which is related to $q^2=(k+k')^2$  via $q^2 = M^2+M'^2-2MM'\omega$, and $s_{13}=(p-k)^2$. Including NP contributions, we have 
\be
{\cal M} = J_{H}^\alpha J^{L}_\alpha+ J_{H} J^{L}+ 
J_{H}^{\alpha\beta} J^{L}_{\alpha\beta}, \\
\ee
with the polarized lepton currents given by 
($u$ and $v$ dimensionful Dirac spinors)
\be
J^{L}_{(\alpha\beta)}(k,k';h) = \frac{1}{\sqrt{8}} 
\bar u_\ell (k') P_h \Gamma_{(\alpha\beta)} 
(1-\gamma_5)v_{\nu_\ell}(k), 
\quad P_h = \frac{1+h\gamma_5\slashed{\widetilde s}}{2}, 
\quad  \Gamma= 1,\, \Gamma_\alpha = \gamma_\alpha,\, 
\Gamma_{\alpha\beta}= \sigma_{\alpha\beta}, \label{eq:lepton-current}
 \ee
where $h=\pm 1$ stands for the two charged lepton helicities, and 
$\widetilde s^\alpha = s^\alpha/m_\ell= (|\vec{k}'|, k^{\prime 0}\hat k')
/m_\ell$ with $\hat k'=\vec{k}'/|\vec{k}'|$ and $m_\ell$ the charged lepton
mass. The $\widetilde s$ polarization vector satisfies the constraints  
$\widetilde s^{\,2}=-1,\, \widetilde s\cdot k'=0$. 

The dimensionless hadron currents read ($c(x)$ and $b(x)$ are Dirac fields 
in coordinate space),
\be
J_{Hrr'}^{(\alpha\beta)}(p,p') =  \langle H_c; p',r'| \bar c(0) 
O_H^{(\alpha\beta)}b(0) | H_b; p, r\rangle, \quad 
 O_H=C_S- C_P \gamma_5,\, O_H^\alpha=\gamma^\alpha(C_V- C_A \gamma_5),
 \, O_H^{\alpha\beta}= C_T\sigma^{\alpha\beta} (1-\gamma_5), \label{eq:Jh}
 \ee
with $C_{V,A}=(1+C_{V_L}\pm C_{V_R})$ and  $C_{S,P}=(C_{S_L}\pm C_{S_R})$. 
The hadron states are normalized as $\langle \vec{p}\,', r'| \vec{p},
r\rangle= (2\pi)^3(E/M)\delta^3(\vec{p}-\vec{p}\,')\delta_{rr'}$, with $r,r'$ 
 spin indexes.

The  lepton tensors needed to obtain  $|{\cal M}|^2$ are 
readily evaluated and they are collected in Appendix~\ref{sec:appL}. 

\subsubsection{Hadron matrix elements}
\label{sec:hme}
 After summing over polarizations, the hadron contributions can be expressed 
in terms of  Lorentz scalar SFs, which depend on $q^2$, the hadron masses and the NP 
Wilson coefficients. To limit their number, it is useful to apply relations 
deduced from Lorentz, parity $({\cal P}) $ and time-reversal  $({\cal T})$ 
transformations of the hadron currents (Eq.~\eqref{eq:Jh}) and states
 ~\cite{Itzykson:1980rh}. Finally, we have ended up with a total of 16 
 independent SFs. 

We illustrate the procedure by discussing in detail here the diagonal 
$J_{H}^\alpha [J_{H}^\rho]^*$ case. The rest of the hadron tensors are compiled in 
Appendix~\ref{sec:appH}, where the technically involved tensor-tensor 
$J_{H}^{\alpha\beta} [J_{H}^{\rho\lambda}]^*$ term is also discussed in  detail. 

The spin-averaged squared of the $O_H^\alpha$ operator matrix element gives 
rise to a (pseudo-)tensor of two indices
\be
W^{\alpha\rho}(p,q, C_V, C_A) = \overline{\sum_{r,r'}} \langle H_c;  p',r' 
|(C_V V^\alpha - C_A A^\alpha)  | H_b; p,r\rangle 
 \langle H_c; p',r' | (C_V V^\rho - C_A A^\rho)| H_b; p,r \rangle^* , 
\ee
with $(C_V V^\alpha - C_A A^\alpha) = \bar c(0) \gamma^\alpha (C_V-C_A\gamma_5) b(0)$. The sum is done over initial (averaged) and final hadron
helicities, and the above tensor should be contracted with  the lepton one 
$L_{\alpha\rho}(k,k';h)$ (Eq.~\eqref{eq:leptensorVA}) to get the contribution
 to $\overline\sum\,  |{\cal M}|^2$. From the above definition, it trivially follows $W^{\alpha\rho}=W^{\rho\alpha*}$ and therefore splitting  $W^{\alpha\rho}$ 
\be
 W^{\alpha\rho}=\frac12[ W^{\alpha\rho}+
 W^{\rho\alpha}]+\frac12[W^{\alpha\rho}-
 W^{\rho\alpha}]\equiv W^{\alpha\rho}_{(s)}
 +W^{\alpha\rho}_{(a)} = \frac12[ W^{\alpha\rho}+
 W^{\alpha\rho*}]+\frac12[W^{\alpha\rho}-
 W^{\alpha\rho*}]\label{eq:defW}
\ee 
we show that the symmetric and antisymmetric parts of the tensor  are real and purely imaginary, respectively. On the other hand, using the time-reversal transformation, 
 we have ($\tilde p^\mu= (p^0, -\vec{p}\,)$)
\bea
W^{\alpha\rho}(p,q, C_V, C_A) &=& \overline{\sum_{r,r'}} \langle H_c;  p',r' 
|{\cal T}^\dagger {\cal T}(C_V V^\alpha - C_A A^\alpha){\cal T}^\dagger 
{\cal T}  | H_b; p,r\rangle 
 \langle H_c; p',r' | {\cal T}^\dagger {\cal T} (C_V V^\rho - C_A A^\rho) 
 {\cal T}^\dagger {\cal T}| H_b; p,r \rangle^* \nonumber \\
 &=&\overline{\sum_{r,r'}} \langle H_c;  \tilde p',r' |C_V^* V_\alpha 
 - C_A^* A_\alpha  | H_b; \tilde p,r\rangle^* 
 \langle H_c; \tilde p',r' | C_V^* V_\rho - C_A^* A_\rho| H_b; \tilde p,r 
 \rangle \nonumber \\
 &=& W^*_{\alpha\rho}(\tilde p,\tilde q, C_V^*, C_A^*). \label{eq:invT}
\eea
Introducing the self-explanatory decomposition,
\be
W^{\alpha\rho}(p,q, C_V, C_A) = |C_V|^2 W^{\alpha\rho}_{VV}(p,q) + 
|C_A|^2 W^{\alpha\rho}_{AA}(p,q) - 
C_V C^*_A W^{\alpha\rho}_{VA} (p,q)- C_A C^*_V W^{\alpha\rho}_{AV}(p,q),
\ee
and using Eq.~\eqref{eq:invT} and the transformation properties under 
parity, we find
\bea
W_{VV}^{\alpha\rho*}(p,q)\stackrel{\cal T}= 
W_{VV\alpha\rho}(\tilde p,\tilde q) 
\stackrel{\cal P}= W_{VV}^{\alpha\rho}(p,q)&,& 
\qquad W_{AA}^{\alpha\rho*}(p,q)\stackrel{\cal T}= 
W_{AA\alpha\rho}(\tilde p,\tilde q) \stackrel{\cal P}= 
W_{AA}^{\alpha\rho}(p,q), \label{eq:PT1}\\
W_{V\!\!A}^{\alpha\rho*}(p,q)\stackrel{\cal T}= 
W_{V\!\!A\alpha\rho}(\tilde p,\tilde q) \stackrel{\cal P}= 
-W_{V\!\!A}^{\alpha\rho}(p,q)&,& \qquad W_{AV}^{\alpha\rho*}(p,q)
\stackrel{\cal T}= W_{AV\alpha\rho}(\tilde p,\tilde q)
 \stackrel{\cal P}= -W_{AV}^{\alpha\rho}(p,q), \label{eq:PT2}
\eea
The above results, along with\footnote{Note that 
by construction $W_{VV,AA}^{\alpha\rho}=W_{VV,AA}^{\rho\alpha*}$ and $W_{AV}^{\alpha\rho}=W_{VA}^{\rho\alpha*}$.} Eq~\eqref{eq:defW}, allows us to conclude  that $W_{VV}^{\alpha\rho}$ and $W_{AA}^{\alpha\rho}$ 
($W_{VA}^{\alpha\rho}$ and $W_{AV}^{\alpha\rho}$) are real symmetric 
tensors (imaginary antisymmetric pseudotensors proportional to the Levi-Civita symbol), and 
$W_{AV}^{\alpha\rho}= W_{VA}^{\alpha\rho}$. The pseudo character of the imaginary tensor is deduced from the behaviour under a parity transformation.  
Therefore, the most general
  expression for $W^{\alpha\rho}(p,q, C_V, C_A)$  reads
\begin{eqnarray}
W^{\alpha\rho}(p,q, C_V,C_A)&=&|C_V|^2 W^{\alpha\rho}_{VV}(p,q) + 
|C_A|^2 W^{\alpha\rho}_{AA}(p,q) - 
2 {\rm Re}(C_V C_A^*) W^{\alpha\rho}_{VA} (p,q)\nonumber \\
&=&-g^{\alpha\rho}
\widetilde W_1+\frac{p^{\alpha}p^{\rho}}{M^2}
\widetilde W_2+i\epsilon^{\alpha\rho\delta\eta}p_{\delta}q_{\eta}
\frac{\widetilde W_3}{2M^2}+\frac{q^{\alpha}q^{\rho}}{M^2}
\widetilde W_4+\dfrac{p^{\alpha}q^{\rho}+p^{\rho}q^{\alpha}}{2M^2}
\widetilde W_5, \nonumber \\ 
\nonumber\\
\widetilde W_{1,2,4,5}(q^2, C_V,C_A) &=& 
|C_V|^2 W_{1,2,4,5}^{VV}(q^2)+ |C_A|^2 W_{1,2,4,5}^{AA}(q^2)\,, \,\, 
\widetilde W_3(q^2,C_V, C_A) = {\rm Re}(C_V C_A^*)W_3^{V\!\!A}(q^2),
\label{eq:wmunu-dec}
\end{eqnarray}
where all $\widetilde W_i$ SFs are real, and  we have used an obvious 
notation in which $W_{1,2,4,5}^{VV}$, $W_{1,2,4,5}^{AA}$ and $W_3^{V\!\!A}$ 
should be obtained from the $W^{\alpha\rho}_{VV}$, $W^{\alpha\rho}_{AA}$ and $-2W^{\alpha\rho}_{VA}$ (pseudo-)tensors. This 
result was previously obtained in  Ref.~\cite{Penalva:2019rgt} for real 
Wilson coefficients.

\subsubsection{ CM and LAB differential decay widths  for an unpolarized final
 charged lepton }
We consider first the case of an unpolarized final charged lepton. 
From the general structure of the lepton and hadron tensors considered in 
this work, which are at most quadratic in $k,k'$, and  in $p$,
respectively, and
using the information on the scalar products compiled in 
Appendix~\ref{sec:scalaP}, one can write the general  expression
\bea
\frac{2\,\overline\sum\,  |{\cal M}|^2}{M^2}\Big|_{\rm unpolarized}= 
{\cal A}(\omega)+{\cal B}(\omega)
 \frac{p\cdot k}{M^2}+ {\cal C}(\omega) \frac{(p\cdot k)^2}{M^4},
 \label{eq:m2abc}
\eea
that is suited to obtain the CM $d^2\Gamma/(d\omega d\cos\theta_\ell)$ and 
LAB $d^2\Gamma/
(d\omega dE_\ell)$ distributions. As already pointed out,   $\theta_\ell$  is the angle made 
by the three-momenta of  the charged lepton and the final hadron in the $W^-$ 
CM system and 
$E_\ell$  the charged lepton energy in the decaying hadron rest frame. The ${\cal A}, {\cal B}$ and ${\cal C}$ functions are linear 
combinations of  the  $\widetilde W$ SFs, introduced in 
Sec.~\ref{sec:hme} and  Appendix~\ref{sec:appH},
and they  depend on $\omega$ as well as  on the lepton and hadron masses. Their expressions 
in terms of the hadronic $\widetilde W$ SFs are given in 
Appendix~\ref{app:coeff}.  As  shown in Subsec.~\ref{sec:hme} and  Appendix~\ref{sec:appH}, the $\widetilde W$
SFs depend on the, generally  complex, Wilson coefficients and  the real SFs ($W'$s) that parameterize the 
 hadron tensors (see an example in Eq.~\eqref{eq:wmunu-dec}).
From Eqs.~\eqref{eq:defsec} and \eqref{eq:m2abc}, taking into account that 
\be
d\omega\, ds_{13}= M M' \left(1-\frac{m^2_\ell}{q^2}\right)
\sqrt{\omega^2-1}\,d\omega\, d\cos\theta_\ell = 2 M d\omega\, dE_{\ell},
\ee
and using the relations in Appendix~\ref{sec:scalaP}  one  gets
\begin{eqnarray}
  \frac{d^2\Gamma}{d\omega d\cos\theta_\ell}& =& \frac{G^2_F|V_{cb}|^2
    M'^3M^2}{16\pi^3}
  \sqrt{\omega^2-1}\left(1-\frac{m_\ell^2}{q^2}\right)^2
   A(\omega,\theta_\ell),
\end{eqnarray}
with
 \begin{eqnarray}
  A(\omega,\theta_\ell) &=& \frac{2\overline\sum\,  
  |{\cal M}|^2}{M^2 \left(1-\frac{m_\ell^2}{q^2}\right)}\Big|_{\rm unpolarized}
  =  a_0(\omega)  + a_1(\omega) \cos\theta_\ell +
  a_2(\omega) \cos^2\theta_\ell,  \label{eq:distr-ang}\label{eq:A}\\ 
  a_0(\omega) &=&\frac{q^2}{q^2-m_\ell^2}\,{\cal A}(\omega)+\frac{M_\omega}{2M}
  \,{\cal B}(\omega)
  +\frac{(q^2-m_\ell^2)M_\omega^2}{4q^2M^2}\,{\cal C}(\omega),
  \nonumber \\
  a_1(\omega) &=& \sqrt{\omega^2-1}\,\frac{M'}{M}\left(\frac{{\cal B}(\omega)}2
  +\frac{(q^2-m_\ell^2)M_\omega}{2q^2M}{\cal C}(\omega)\right), \nonumber \\
  a_2(\omega)&=& (\omega^2-1)\frac{M'^2}{M^2}\frac{q^2-m_\ell^2}{4q^2}\, 
  {\cal C} (\omega),\label{eq:as}
\end{eqnarray}
and where $M_\omega=M-M'\omega$. For the LAB differential decay with we obtain
\begin{eqnarray}
   \frac{d^2\Gamma}{d\omega dE_\ell}& =& \frac{G^2_F|V_{cb}|^2
    M'^2M^2}{8\pi^3} C(\omega,E_\ell), 
\end{eqnarray}
with
\begin{eqnarray}
 && C(\omega,E_\ell) = \frac{2\overline\sum \, |{\cal M}|^2}{M^2}\Big|_{\rm unpolarized}= 
     c_0(\omega)  + c_1(\omega) \frac{E_\ell}{M} +   c_2(\omega) 
     \frac{E^2_\ell}{M^2}, \label{eq:distr-ene}\label{eq:C}\\
&&  c_0 (\omega) = {\cal A}(\omega)+\frac{M_\omega}{M}\,{\cal B}(\omega)+
 \frac{M_\omega^2}{M^2}\, {\cal C}(\omega), \qquad
  c_1(\omega) =   -{\cal B}(\omega)-\frac{2M_\omega}{M}\,{\cal C}(\omega), 
  \qquad
  c_2(\omega)= {\cal C}(\omega). \label{eq:ces}
\end{eqnarray}
 The variable $\omega$
varies from 1 to $\omega_{\rm max}= (M^2+M'^2-m^2_{\ell})/(2MM')$ and
$\cos\theta_\ell$ between $-1$ and 1, while $E_\ell\in[E_\ell^{-},E_\ell^{+}]$, where 
\begin{equation}
  E_\ell^{\pm}= \frac{(M-M'\omega)(q^2+m^2_\ell)  
  \pm M' \sqrt{\omega^2-1}(q^2-{m^2_\ell})}{2q^2}.
\end{equation}  

The first result of this work is that the inclusion of NP contributions  does 
not induce further terms in the $\cos\theta_\ell$ and $E_\ell$ expansions 
of $A(\omega,\theta_\ell)$ and $C(\omega,E_\ell)$ with respect to a pure SM
calculation. This result was  already  obtained in 
 Ref.~\cite{Penalva:2019rgt} although, there, the effects of the tensor 
 ${\cal O}_{T}$ NP term and of 
 complex Wilson coefficients were neglected. From Eqs.~\eqref{eq:as} and 
 \eqref{eq:ces}, which derive
directly from the general expression in Eq.~\eqref{eq:m2abc}, one now clearly 
understands  that the universal
function $\frac{M^2}{M^{\prime\,2}}\frac{a_2(\omega)}
{(1-m_\ell^2/q^2)c_2(\omega)}=(\omega^2-1)/4$, that we discussed 
in Ref.~\cite{Penalva:2019rgt},
has in fact a purely kinematical origin and
it should be obtained in any physics scenario in which the lepton tensors 
are at
most quadratic in the lepton momenta. 
We  stress here again  that, 
although the effective Hamiltonian in Eq.~\eqref{eq:hnp} refers to $b\to c$ 
transitions, all expressions are general and apply independently of the 
quark flavors involved
 in the NP four-fermion operators.

Focusing on the LAB distribution, we see that $c_2(\omega)$ determines 
${\cal C}(\omega)$, and the latter 
together with $c_1(\omega)$ fixes the function ${\cal B}(\omega)$. Finally, 
${\cal A}
(\omega)$ is obtained from $c_0(\omega),\,{\cal B}(\omega)$ and 
${\cal C}(\omega)$. 
The discussion is totally similar for the CM angular differential decay width. 
Indeed, 
the  unpolarized $d^2\Gamma/(d\omega d\cos\theta_\ell)$ and 
$d^2\Gamma/(d\omega dE_\ell)$ distributions turn out to be equivalent in the
sense that  both of 
them provide the same information on the Hamiltonian which induces the 
semileptonic  decay: three different linear combinations of the $\widetilde W$ SFs. 
 Additional  information can be obtained by  considering the dependence on $m_\ell$ of the unpolarized decay 
 distributions and using simultaneously data for  the $\tau$ and 
 $\ell=e$ or $\mu$ (massless in good approximation) decay modes. Indeed, up to a total
  of five linear combinations of SFs can be determined, since ${\cal C}(\omega)$
   does not depend on $m_\ell$. For instance in the SM, the massless decay fixes 
   $W_{1,2,3}$, while $W_4$ and $W_5$ can be obtained 
from  the tau mode ${\cal A}(\omega)$ and ${\cal B}(\omega)$ functions, respectively.
  Thus,  for $q^2\ge m_\tau^2$,
all SFs can be  determined from unpolarized distributions when NP is not present. 
This implies that for a final $\tau$ lepton the SM  
CM $d^2\Gamma/(d\omega d\cos\theta_\tau)$ and LAB $d^2\Gamma/(d\omega dE_\tau)$ 
polarized distributions
 can be determined from  unpolarized $\mu, e$ 
and $\tau$ data alone. In our previous work in Ref.~\cite{Penalva:2019rgt}, 
we wrongly concluded that this was not possible for the latter distribution.

\subsubsection{ CM and LAB differential decay widths  for a polarized final charged lepton }

For a polarized final charged lepton,  in reference systems, like the CM 
and the LAB ones
considered in this work, for which 
$\epsilon_{\delta\eta\mu\nu}k^\delta q^\eta s^\mu p^\nu=0$, and for 
the contributions we have, one can generally write
\begin{eqnarray}
\frac{2\,\overline\sum\,  |{\cal M}|^2}{M^2} &= & \frac12
\left[{\cal A}(\omega)
+{\cal B}(\omega) \frac{(p\cdot k)}{M^2}+ {\cal C}(\omega) 
\frac{(p\cdot k)^2}{M^4}
\right]\nonumber \\
&&+ h\left[\frac{(p\cdot s)}{Mm_\ell}\left({\cal A_H}(\omega)
+ {\cal C_H}(\omega)
 \frac{(k\cdot p)}{M^2}\right) +\frac{(q\cdot s)}{Mm_\ell}\left( 
 {\cal B_H}(\omega)
  + {\cal D_H}(\omega) \frac{(k\cdot p)}{M^2}+ {\cal E_H}(\omega) 
  \frac{(k\cdot p)^2}
  {M^4}\right)\right],\label{eq:pol}
\end{eqnarray}
where five new independent functions ${\cal A_H}, {\cal B_H}, 
{\cal C_H}, {\cal D_H}$ and
 ${\cal E_H}$ are now needed. They can be written in terms of the 
 $\widetilde W$ SFs and the
 corresponding expressions  are given in 
 Appendix~\ref{app:coeff}.  Note that 
 Eq.(\ref{eq:pol}) does not  diverge in the
 $m_\ell\to 0$ limit. Since this $1/m_\ell$ dependence originates from the 
 $P_h$
 projector present in the lepton
 tensor, the easiest way to find the $m_\ell\to 0$ leading behaviour  is by looking at the general lepton 
 tensor expression in Eq.~(\ref{eq:LT}) and realizing that the factor
  $P_h(\slashed{k}'+ m_{\ell})=(\slashed{k}'+ m_{\ell})P_h$ reduces to 
  $\left(\slashed{k}'(1-h\gamma_5)+{\cal O}(m_\ell)\right)$ in that limit.
This result, together with Eq.~(\ref{eq:LT}), also tell us  that for a 
massless 
charged lepton, the $h=+1$ lepton tensors vanish, as expected from 
conservation of chirality, except for those corresponding 
to the diagonal and interference ${\cal O}_{S_{L,R}}$ and ${\cal O}_T$ NP 
operators. On the other hand, for $h=-1$, and in the massless limit, only 
  the lepton tensor originating from the diagonal ${\cal O}_{V_{L,R}}$ terms 
  are nonzero.

 For this polarized case one  finds that
 \bea
 A_h(\omega,\cos\theta_\ell)= \frac{2\overline\sum\,  |{\cal M}|^2}
 {M^2 \left(1-\frac{m_\ell^2}{q^2}\right)}=a_0(\omega,h)+a_1(\omega,h)
 \cos\theta_\ell+
 a_2(\omega,h)\cos^2\theta_\ell,
 \eea
 where now
 \bea
  a_0(\omega,h)&=&\frac12\left( a_0(\omega)+h
 \left\{\frac{M}{m_\ell}\left[\frac{M_\omega}{M}{\cal A_H} 
 +\frac{M^2_\omega}{2M^2}{\cal C_H} + \frac{q^2}{M^2}\left ({\cal B_H}
 + \frac{M_\omega}{2M} {\cal D_H} + \frac{M_\omega^2}{4M^2} {\cal E_H}
 \right)\right]\right.\right.
 \nonumber \\
&&\hspace{2.15cm}\left.\left.-\frac{m_\ell}{M}\frac{M_\omega}{2M}
\left[{\cal D_H} 
+\frac{M_\omega}{M}\left( {\cal E_H} +\frac{M^2}{q^2} {\cal C_H}\right)\right] 
+ \frac{m^3_\ell M_\omega^2}{4M^3q^2}{\cal E_H}\right\}\right),\nonumber \\
 a_1(\omega,h)&=&\frac12\left( a_1(\omega)+h\frac{M'}{M}\sqrt{\omega^2-1}
 \Bigg\{\frac{M}{m_\ell}\left[\frac{m_\ell^2+q^2}{m_\ell^2-q^2}{\cal A_H}+ 
 \frac{q^2}{2M^2}\left({\cal D_H}+\frac{M_\omega}{M} {\cal
 E_H}\right)\right]\right.\nonumber \\
&&\hspace{4.15cm}\left. -\frac{m_\ell}{M}
\left[\frac{{\cal D_H}}{2}+ \frac{M_\omega}{M}\left( {\cal E_H} 
+\frac{M^2}{q^2} {\cal C_H}\right)\right]+ \frac{m^3_\ell M_\omega}{2M^2q^2}
{\cal E_H}\Bigg\}\right),\nonumber \\
 a_2(\omega,h)&=&\frac12\left( a_2(\omega)+h \frac{M^{\prime 2}}{M^2}
 \left(\omega^2-1 \right)\left\{\frac{M}{4m_\ell}\left( \frac{q^2}{M^2}
  {\cal E_H} -2 {\cal C_H}\right)-
\frac{m_\ell}{2M}\left({\cal E_H} +  \frac{M^2}{q^2}{\cal C_H}\right)
+ \frac{m^3_\ell}{4Mq^2}{\cal E_H}\right\}\
\right)\label{eq:aspol}
 \eea
 For the LAB distribution, the decomposition into $h=\pm 1$ contributions 
 is more involved  and we find\footnote{Note that the definition of the coefficients 
$\widehat{c}_{0,1}$ given in Ref.~\cite{Penalva:2019rgt} differ from that 
adopted here by a factor $m^2_\ell/M^2$.}
\begin{eqnarray}
 C_{h}(\omega,E_\ell)= \frac{2\overline\sum  |{\cal M}|^2}
 {M^2 }= \frac{C(\omega,E_\ell)}{2}-\frac{h}{2}\frac{M}{p_\ell}
 \left(\widehat{c}_0+
     \left[c_0+\widehat{c}_1\right] \frac{E_\ell}{M}+
     \left[c_1+\widehat{c}_2\right] \frac{E^2_\ell}{M^2}+
    \left[ c_2+\widehat{c}_3\right] \frac{E^3_\ell}{M^3}\right), \label{eq:Chpol}
\end{eqnarray}
with $C(\omega,E_\ell)$ the corresponding unpolarized function introduced in Eq.~(\ref{eq:distr-ene}), 
$p_\ell=(E_\ell^2-m_\ell^2)^{\frac12}$ the charged lepton three-momentum in the LAB system, and 
\begin{eqnarray}
\widehat{c}_0(\omega) & = & \frac{2m_\ell}{M}\left( {\cal A_H} 
+\frac{M_\omega}{M} \left({\cal B_H}+ {\cal C_H}\right) 
+ \frac{M_\omega^2}{M^2} {\cal D_H}
 + \frac{M_\omega^3}{M^3} {\cal E_H} \right ), \nonumber \\
 \widehat{c}_1(\omega) & = & - c_0 -\frac{M}{m_\ell}\frac{q^2}{M^2}
 \left({\cal B_H} +  \frac{M_\omega}{M} {\cal D_H}+ 
 \frac{M_\omega^2}{M^2} {\cal E_H} \right)-\frac{m_\ell}{M}
 \left({\cal B_H} +2 {\cal C_H} + \frac{3M_\omega}{M}{\cal D_H}  
 + \frac{5M_\omega^2}{M^2}{\cal E_H}\right), \nonumber \\
 \widehat{c}_2(\omega) & = & - c_1  + \frac{M}{m_\ell}
 \left[\frac{q^2}{M^2}\left( {\cal D_H} 
 + \frac{2M_\omega}{M} {\cal E_H}\right) 
  -2{\cal A_H} -\frac{2M_\omega}{M}{\cal C_H} \right]
  + \frac{m_\ell}{M}\left({\cal D_H}+ \frac{4M_\omega}{M} 
  {\cal E_H}\right), \nonumber \\
 \widehat{c}_3(\omega) & = & - c_2  + \frac{M}{m_\ell}
 \left(2\,{\cal C_H}-\frac{q^2}{M^2} {\cal E_H} \right)-
 \frac{m_\ell}{M} {\cal E_H}.
\label{eq:cespol}
\end{eqnarray} 
 From the discussion above, we know that the coefficients of the 
 $M/m_\ell$ terms
 in Eqs.~\eqref{eq:aspol} and \eqref{eq:cespol} should  vanish at least 
 as ${\cal O}(m_\ell)$ in the $m_\ell\to0$ limit, 
 guaranteeing that
 the differential decay widths are finite in that limit.
 
  Unlike the unpolarized case, where 
 ${\cal A}, {\cal B}$ and ${\cal C}$ could be determined either from the CM 
 or LAB distributions, both LAB and CM 
 helicity distributions are now needed simultaneously to obtain all 
 five ${\cal A_H}, {\cal B_H}, {\cal C_H}, {\cal D_H}$ and
 ${\cal E_H}$ additional functions. This is so since in this case only four of them can be determined from the $E_\ell$ dependence of the polarized
 $d^2\Gamma/(d\omega dE_\ell)$ distribution, while the use of the 
 $\cos\theta_\ell$ dependence of the polarized $d^2\Gamma/(d\omega 
 d\cos\theta_\ell)$ distribution only gives access to three of them.
 
In this polarized case, even assuming that the NP terms affect
 only to the third lepton family, the strategy 
to obtain polarized information for $\tau$ decays from non-polarized data 
is spoiled by the presence of the ${\cal O}_{T} $ NP operators. This is
 due to the diagonal 
$O_H^{\alpha\beta} O_H^{\rho\lambda*}$ and the  
$O^\alpha_HO_H^{\rho\lambda *}$ interference terms. We can, for example, 
better understand this by
looking at Eqs.~\eqref{eq:aesplus1} and \eqref{eq:aesmenos1}, where we give 
the  coefficients $a_{0,1,2}(\omega,h)$    directly in terms of the $\widetilde W$ SFs. There,  
we observe that both $a_{0,1,2}(h=+1)$ and  $a_{0,1,2}(h=-1)$ have contributions proportional to $
m_\ell$. Therefore the 
angular coefficients for $h=-1$ 
cannot be measured in the charged lepton massless decays, since such reactions
 do not provide information about the $\widetilde W^T_{2,3,4}$ and 
 $\widetilde W_{I4,I6,I7}$ contributions to $a_{0,1,2}(h=-1)$. This remains true
  even if NP 
 existed in the first and second generations. Therefore, the $h=+1$ and 
 $h=-1$ parts cannot be disentangled from the measurement of the unpolarized  
 $d^2\Gamma/(d\omega d\cos\theta_{\ell=e,\mu,\tau})$ distributions alone, although 
 non-polarized $d^2\Gamma/(d\omega d\cos\theta_{\ell=e,\mu})$ data can be used. 

We would also note that the ${\cal O}_{S_{L,R}}$ and ${\cal O}_{T} $ NP operators lead to non-vanishing contributions ($\widetilde W_{SP}$, $\widetilde W_{2,3,4}^T$ and $\widetilde W_{I3}$) for positive helicity in the massless charge lepton limit.

The discussion is similar for the LAB $d^2\Gamma/(d\omega dE_\tau)$ differential decay width.

\section{Semileptonic $\Lambda_b^0\to  \Lambda_c^+\ell^-\bar\nu_\ell$ decay}
\label{sec:LbtoLc}
We apply here the  general formalism derived in the previous sections to 
the study of the  semileptonic $\Lambda_b\to\Lambda_c $ decay, paying attention to the 
NP corrections to the SM results. We update the theoretical framework and the numerical results presented previously in Ref.~\cite{Penalva:2019rgt}, where NP tensor terms were not considered and the  Wilson coefficients were 
taken to be real. 
We have used  the LQCD form-factors derived in  Refs.~\cite{Detmold:2015aaa, Datta:2017aue} and the best-fit Wilson coefficients determined in \cite{Murgui:2019czp}. We anticipate that  this new  comprehensive analysis confirms most of the 
findings of Ref.~\cite{Penalva:2019rgt}, and  shows that
the  double differential LAB $d^2\Gamma/(d\omega dE_\ell)$ or CM $d^2\Gamma/(d\omega 
d\cos\theta_\ell$) distributions of this decay can be used to  distinguish 
between different  NP fits to $b\to c\tau\bar\nu_\tau$ anomalies in the meson sector,  that otherwise give the same  total and differential $d\Gamma/d\omega$ widths.

\subsection{Form Factors and SFs}
The relevant hadronic  matrix
elements  can be parameterized in terms of one scalar ($F_S$), one 
pseudo-scalar ($F_P$),  three vector ($F_i$),  three axial
($G_i$) and four tensor ($T_i$) form-factors, which are real functions 
of $\omega$ and that are greatly
constrained by HQSS near zero recoil 
($\omega=1$)~\cite{Neubert:1993mb,Bernlochner:2018bfn,Bernlochner:2018kxh}
\begin{eqnarray}
 \langle\Lambda_c;\vec{p}^{\,\prime};r'| \bar c(0)\left(1-\gamma_5\right) b(0)|
 \Lambda_b;\vec{p};r\rangle &=& \bar{u}^{(r')}_{\Lambda_c}
 (\vec{p}^{\,\prime}\,)\left(F_S-\gamma_5 F_P\right) 
 u^{(r)}_{\Lambda_b}(\vec{p}\,), \nonumber \\
\langle\Lambda_c;\vec{p}^{\,\prime};r'|  V^\alpha-A^\alpha  
|\Lambda_b;\vec{p};r\rangle
&=& 
\bar{u}^{(r')}_{\Lambda_c}(\vec{p}^{\,\prime}\,)\Bigg\{ \gamma^\alpha 
\left(F_1-\gamma_5G_1\right)  + \frac{p^{\alpha}}{M_{\Lambda_b}} 
\left(F_2-\gamma_5G_2\right) +
 \frac{p^{\prime\alpha}}{M_{\Lambda_c}}\left(F_3-\gamma_5G_3\right) \Bigg\} 
 u^{(r)}_{\Lambda_b}(\vec{p}\,) \nonumber \\
\langle\Lambda_c;\vec{p}^{\,\prime};r'|  \bar c(0)\sigma^{\alpha\beta}b(0)   
|\Lambda_b;\vec{p};r\rangle &=& 
 \bar{u}^{(r')}_{\Lambda_c}(\vec{p}^{\,\prime}\,)
 \Bigg\{i\frac{T_1}{M^2_{\Lambda_b}}
 \left(p^\alpha p^{\prime \beta}-p^\beta p^{\prime \alpha}\right) 
 + i\frac{T_2}{M_{\Lambda_b}}\left(\gamma^\alpha p^\beta
  -\gamma^\beta p^\alpha  \right) \nonumber \\
&&\hspace{1.5cm}+ i\frac{T_3}{M_{\Lambda_b}}\left(\gamma^\alpha p^{\prime \beta}
-\gamma^\beta p^{\prime \alpha}\right) 
+ T_4 \sigma^{\alpha\beta} \Bigg\} u^{(r)}_{\Lambda_b}(\vec{p}\,),
\label{eq.FactoresForma}
\end{eqnarray}
with  $u_{\Lambda_b, \Lambda_c}$ dimensionless Dirac spinors (note that 
for leptons we use spinors with square root mass dimensions instead). In the
heavy quark limit all the above form factors either vanish or equal
the leading-order Isgur-Wise function~\cite{Bernlochner:2018bfn} $\zeta(\omega)$, satisfying $\zeta(1)=1$
\begin{equation}
F_2= F_3= G_2 = G_3=T_1=T_2=T_3 =0, \quad  F_1(\omega) = G_1(\omega) = F_S(\omega)= F_P(\omega) = T_4(\omega) =\zeta(\omega)\label{eq:hqss}
\end{equation}
Moreover, as discussed  in \cite{Neubert:1993mb, Bernlochner:2018bfn} no additional unknown functions beyond $\zeta(\omega)$ are
needed to parametrize the ${\cal O}(\Lambda_{\rm QCD}/m_{b,c})$ corrections.
Perturbative corrections to the heavy quark currents can
be computed by matching QCD onto heavy quark effective theory and
introduce no new hadronic parameters. The same also holds for the order ${\cal O}(\alpha\Lambda_{\rm QCD}/m_{b,c})$.

The hadron tensors are readily obtained using 
\be
 \overline{\sum_{r,r'}} \langle \Lambda_c;  p',r' | \bar c(0) 
 \Gamma^{(\alpha\beta)} b(0) | \Lambda_b; p,r\rangle 
 \langle \Lambda_c; p',r' | \bar c(0) \Gamma^{(\rho\lambda)} b(0) 
 | H_b; p,r \rangle^* 
= \frac12 {\rm Tr}\Big[ \frac{\slashed{p}'+M_{\Lambda_c}}{2M_{\Lambda_c}}
F_{\Gamma}^{(\alpha\beta)} \frac{\slashed{p}+M_{\Lambda_b}}{2M_{\Lambda_b}}
\gamma^0 F_{\Gamma}^{(\rho\lambda)\dagger}\gamma^0 \Big], \label{eq:wmunu-baryons}
\ee
with the Dirac matrices 
\begin{eqnarray}
F_{\Gamma}^{(\alpha\beta)}&=& 1,\, \gamma_5, \, \left(\gamma^\alpha F_1 
+ \frac{p^\alpha}{M_{\Lambda_b}} F_2 +
 \frac{p^{\prime\alpha}}{M_{\Lambda_c}} F_3\right), \, \left(\gamma^\alpha\gamma_5 G_1 + \frac{p^\alpha}{M_{\Lambda_b}}\gamma_5 G_2 +
 \frac{p^{\prime\alpha}}{M_{\Lambda_c}}\gamma_5 G_3\right), \nonumber \\
 &&  \left[i\frac{T_1}{M^2_{\Lambda_b}}\left(p^\alpha p^{\prime \beta}-p^\beta p^{\prime \alpha}\right)+ i\frac{T_2}{M_{\Lambda_b}}\left(\gamma^\alpha p^\beta -\gamma^\beta p^\alpha  \right)+ i\frac{T_3}{M_{\Lambda_b}}\left(\gamma^\alpha p^{\prime \beta}-\gamma^\beta p^{\prime \alpha}\right) + T_4 \sigma^{\alpha\beta}\right], \nonumber \\
 &&\epsilon^{\alpha\beta\delta\eta}\left[T_1 \frac{p_\delta p^\prime_\eta}{M^2_{\Lambda_b}}+ T_2 \gamma_\delta \frac{p_\eta}{M_{\Lambda_b}} + T_3 \gamma_\delta \frac{p^\prime_\eta}{M_{\Lambda_b}} 
 + \frac12 T_4 \gamma_\delta\gamma_\eta\right]. \label{eq:FG}
 \end{eqnarray}
The last of the structures in  Eq.~\eqref{eq:FG} accounts for the matrix
 element of the operator $\bar c(0)\sigma^{\alpha\beta}\gamma_5 b(0)$ 
 between the initial and final hadrons which, thanks to Eq.~\eqref{eq:sigma}, 
 is related to that of the tensor operator $\bar c(0)\sigma^{\alpha\beta} b(0)$.

From Eq.~\eqref{eq:wmunu-baryons} one  can obtain the $\widetilde W$ SFs, 
and hence the LAB  $d^2\Gamma/(d\omega dE_\ell)$ and 
CM $d^2\Gamma/(d\omega d\cos\theta_\ell)$ distributions, in terms of the 
Wilson coefficients and form-factors introduced  in  
Eqs.~\eqref{eq:hnp} and \eqref{eq.FactoresForma}, respectively. The explicit expressions are given in  
Appendix~\ref{app:ff}.  As detailed also in this appendix,  
the form factors used in Eq.~\eqref{eq.FactoresForma} are easily related 
to those computed in 
the LQCD simulations of Refs.~\cite{Detmold:2015aaa} (vector and axial) and 
\cite{Datta:2017aue} (tensor), which were given in terms of the 
Bourrely-Caprini-Lellouch parametrization~\cite{Bourrely:2008za} (see Eq.~(79) 
of \cite{Detmold:2015aaa}). On the other hand, the scalar ($F_S$) and 
pseudoscalar ($F_P$) form factors  
are directly related (see  Eqs.~(2.12) and (2.13) of Ref.~\cite{Datta:2017aue}) to the  $f_0$ vector and $g_0$ axial  ones  obtained in the LQCD calculation of  Ref.~\cite{Detmold:2015aaa}.
For  numerical calculations,  we use here for the vector, axial  and 
tensor form-factors, the 11 and 7 parameters given in Table VIII  of 
Ref.~\cite{Detmold:2015aaa}  
and Table 2 of  Ref.~\cite{Datta:2017aue}, respectively. To assess the 
uncertainties of the observables that depend
of the form factors, we have included the (cross) correlations between 
all the parameters of  the ten (vector, axial vector, and tensor) form 
factors, as provided in the supplemental files of Ref.~\cite{Datta:2017aue}.
\subsection{Results: NP effects for $\Lambda_b^0\to  \Lambda_c^+\tau^-
\bar\nu_\tau$ decay}
In this section, we will present numerical results using the Wilson coefficients corresponding to the independent Fits 6 and 7
of Ref.~\cite{Murgui:2019czp}, which are real as corresponds to a scheme where the CP symmetry is preserved. Details of four different fits (4, 5, 6 and 7), that include
all the NP terms given in Eq.~\eqref{eq:hnp}, are provided in the Table 6 of that work. We do not consider the scenarios determined by Fits 4 and 5 because they describe an unlikely physical situation in which the SM coefficient is almost canceled and its effect is replaced by NP contributions. 
The exhaustive  analysis carried out in Ref.~\cite{Murgui:2019czp} is a cutting-edge LFUV study in semileptonic $B\to D^{(*)}$ decays. 
The data used for the fits include  the ${\cal R}_D$ and ${\cal R}_{D^*}$ ratios,
the normalized experimental distributions of $d\Gamma(B\to D \tau \bar\nu_\tau)/dq^2$ and $d\Gamma(B\to D^* \tau \bar\nu_\tau)/dq^2$ measured by Belle and BaBar as well as the  longitudinal polarization fraction 
$F_L^{D^*}=\Gamma_{\lambda_{D^*}=0}(B\to D^{*}\tau \bar \nu_\tau)/\Gamma(B\to D^{*}\tau \bar \nu_\tau)$ provided by Belle. 
The $\chi^2$ merit function is defined in  Eq.~(3.1)  of Ref.~\cite{Murgui:2019czp}, and it is constructed taking into account  
the above data inputs and some prior knowledge of the $B \to D $  and $B \to D^*$ semileptonic form-factors. In addition, 
some upper bounds   on the leptonic decay rate $B_c\to \tau \nu_\tau$ are imposed by allowing only points in the parameter
space that fulfill this bound. 

 Before discussing the results, we dedicate a few words about how we estimate the uncertainties that affect our predictions. We use Monte Carlo error  propagation to maintain, when possible, statistical correlations between the different parameters involved in our calculations. The first source of uncertainties is found in the form factors. This is in fact  the  theoretical error in the case of the results obtained 
 within the  SM. Thus, SM results will be presented with  an error  band that we obtain using the covariance 
matrix provided as supplemental material in Ref.~\cite{Datta:2017aue} and that
accounts  for   68\% confident level (CL) intervals. 
Results including NP contributions are not only affected by the LQCD form-factors errors but also 
 by the uncertainties in the fitted Wilson  coefficients. To evaluate
the latter, for each of Fits 6 and 7, we use different sets of Wilson coefficients 
provided by the authors of Ref.~\cite{Murgui:2019czp}. They have been obtained through 
successive small steps in the multiparameter space, with each step leading to a moderate $\chi^2$ enhancement.
We use  1$\sigma$ sets, values  of the Wilson coefficients for which $\Delta\chi^2\le 1$ with respect to its minimum value, to generate the distribution of each observable, taking into account in this way statistical correlations. From this derived distributions, we determine the maximum 
deviation above and below its central value, the latter
obtained with the values of the Wilson coefficients corresponding to the minimum of $\chi^2$. These deviations define the, asymmetric in general, uncertainty associated with the NP Wilson coefficients.  The latter  uncertainty is then added
in quadratures with the one corresponding to the form factors determination to define an error (asymmetric) band. Thus, results 
obtained including NP will always be provided with such an error band. To get an idea of the relative relevance of both sources of theoretical error,
in many cases, the smallest-in-size bands associated only  with the uncertainties in the form factors will  
 also be shown.
\begin{figure}[tbh]
\includegraphics[height=4.cm]{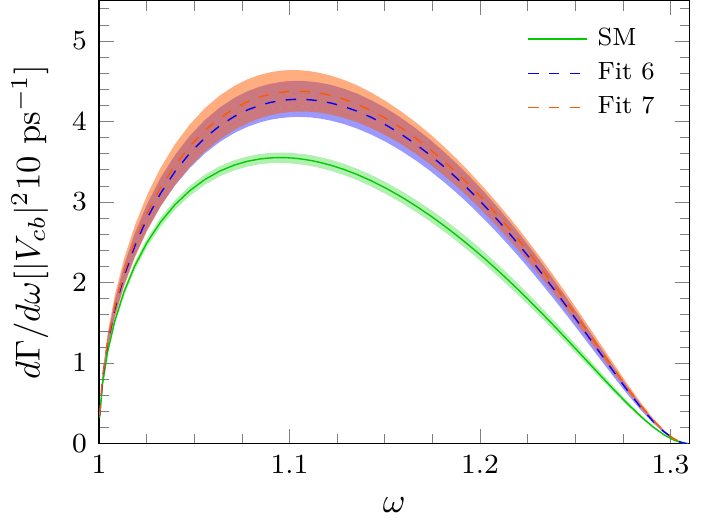} \hspace{.15cm}
\includegraphics[height=4.cm]{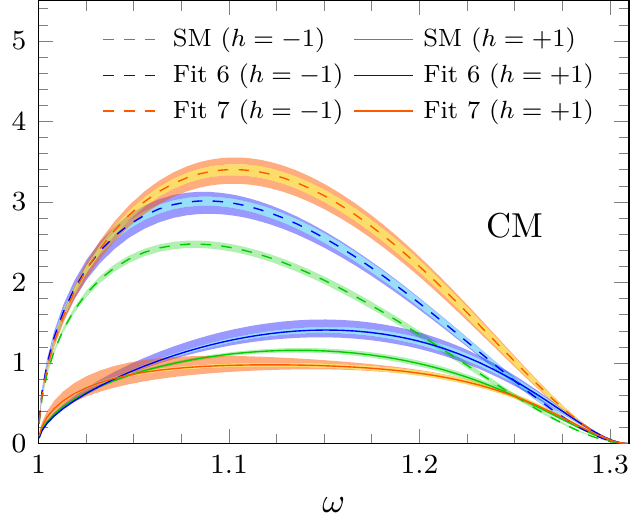}\hspace{.15cm} 
\includegraphics[height=4.cm]{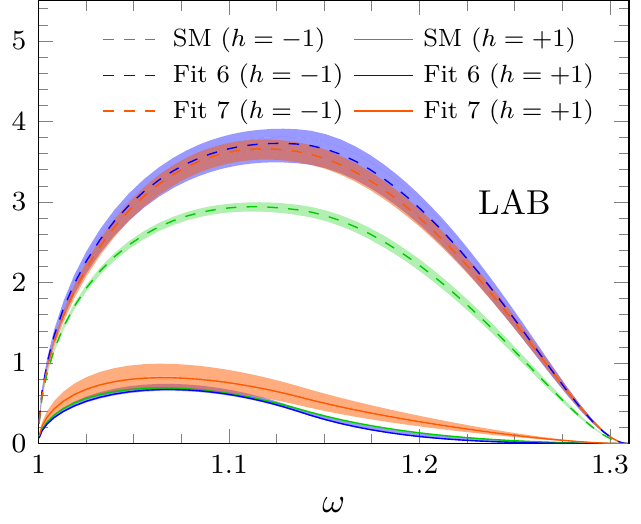} 
\caption{ Left panel: $d\Gamma(\Lambda_b\to\Lambda_c \tau\bar \nu_\tau)/d\omega$ 
differential decay width, as a function of $\omega$ and in units of $10 |V_{cb}|^2 {\rm ps}^{-1}$. We show SM predictions and full NP results obtained including all terms in Eq.~\eqref{eq:hnp} and using the Wilson coefficients from 
Fits 6  and  7  of Ref.~\cite{Murgui:2019czp}. In the middle and right panels, we  show the contributions to 
the $d\Gamma/d\omega$  corresponding to $\tau$ leptons with well defined helicities ($h=\pm 1)$ in the  $W^-$CM and LAB reference systems, respectively.  Uncertainty bands are obtained as detailed in the main text. 
 For the NP results in the CM distributions, we also show the error bands 
 corresponding to the form-factors uncertainties, which can be seen 
 in  lighter colors within the total error bands. In the case of 
 the LAB frame, 
 SM and Fit 6 positive-helicity distributions are practically indistinguishable.}  
\label{fig:dgdw1}
\end{figure} 
\begin{table}
\begin{center}
\begin{tabular}{c|ccc}
                        &  SM & Fit 6 \cite{Murgui:2019czp} & Fit 7 \cite{Murgui:2019czp}\\\hline\tstrut
 $\Gamma(\Lambda_b\to \Lambda_c e(\mu)\bar\nu_{e(\mu)})/\left(10\times |V_{cb}|^2 {\rm ps}^{-1}\right)$ & ~$2.15\pm 0.08$ & $-$ & $-$\\ \tstrut 
 $\Gamma(\Lambda_b\to \Lambda_c \tau\bar\nu_\tau)/ \left(10\times |V_{cb}|^2 {\rm ps}^{-1}\right)$ & ~$0.715^{+0.014}_{-0.016}$ & ~$0.872\pm 0.047$&~  $0.892\pm 0.051$\\ \tstrut
 ${\cal R}_{\Lambda_c}$ & ~$0.332 \pm 0.007$ & ~$0.404\pm 0.022$  & ~ $0.414\pm 0.024$ \\ \hline
\end{tabular}
\end{center}
\caption{Total widths and  ${\cal R}_{\Lambda_c}$ values associated to the distributions shown in the left panel of Fig.~\ref{fig:dgdw1}. }
\label{tab:ratios}
\end{table}
We start by showing in Fig.~\ref{fig:dgdw1} results for the 
$d\Gamma(\Lambda_b\to\Lambda_c\tau\bar\nu_\tau)/d\omega$ differential decay 
width. As we see in the left panel, Fits 6 and 7 give very similar 
 results for $d\Gamma/d\omega$ and  they become
 indistinguishable once the full uncertainty band is taken into account. Thus, by looking at 
 the $d\Gamma/d\omega$ differential 
  decay width (or the integrated decay width for
 that matter, as can be seen
  in Table~\ref{tab:ratios}) 
 one could not decide which fit, and thus what NP terms,
 would be preferable  to explain the data. As compared to our partial results 
 of Ref.~\cite{Penalva:2019rgt}, we find a quite significant 
 reduction of the error bands of the NP distributions thanks to having considered here the statistical correlations between the Wilson coefficients.
 
 In Fig.~\ref{fig:dgdw1}, we also show the 
 separate contributions to $d\Gamma/d\omega$ corresponding to 
    $\tau$ leptons with well defined helicity ($h=\pm1$)  
  measured either in the CM (middle panel) or the LAB (right panel) reference
 systems. In the latter case there is no clear distinction (once the full error band is
 taken into account) between the predictions
 corresponding to Fits 6 and 7. The situation clearly improves
  for the case of well defined helicities in the 
 CM system where the predictions from the two fits can be told apart 
 in most of the $\omega$ range. However, polarized distributions are very challenging measurements because of the presence of 
 undetected neutrinos, so next we examine other possibilities. 

\begin{figure}[tbh]
\includegraphics[height=4.15cm,width=5.6cm]{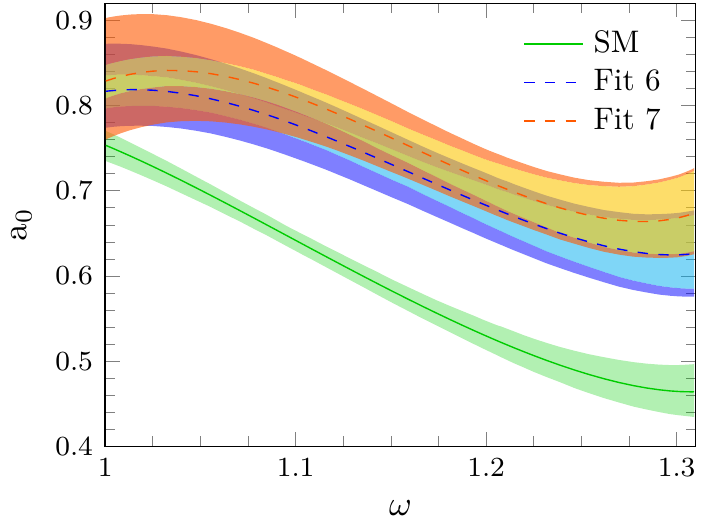}\hspace{.15cm} 
\includegraphics[height=4.15cm,width=5.6cm]{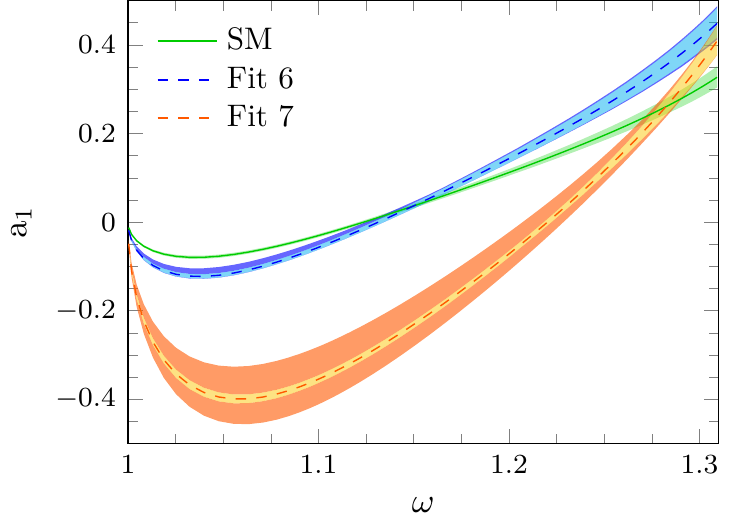} \hspace{.15cm} 
\includegraphics[height=4.29cm,width=5.6cm]{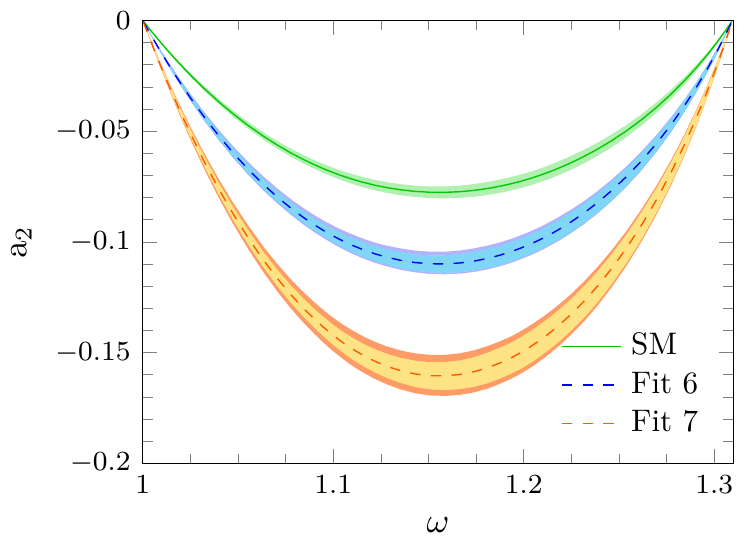}\hspace{.15cm} 
\includegraphics[height=4.29cm,width=5.6cm]{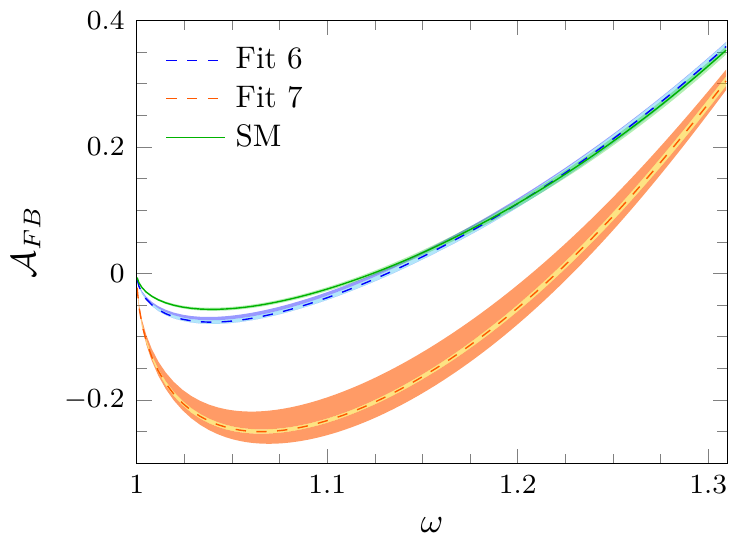}
\caption{ Top: CM angular expansion coefficients $a_0,a_1$ and $a_2$   
for the unpolarized  
$d^2\Gamma[\Lambda_b\to\Lambda_c \tau\bar \nu_\tau]/(d\omega d\cos\theta_\ell)$ 
differential decay width (Eqs.~\eqref{eq:A} and \eqref{eq:as}), as a 
function of $\omega$. Bottom: forward-backward asymmetry, 
${\cal A}_{FB}= a_1/(2a_0+2a_2/3)$. We show SM 
 and full results, the latter evaluated  including all NP terms in 
 Eq.~\eqref{eq:hnp}
 and using the Wilson coefficients from Fits 6 and  7 of 
 Ref.~\cite{Murgui:2019czp}.
 Uncertainty bands are obtained as  explained in the main text. 
 For NP results we also show the error bands 
  corresponding to the form-factors uncertainties, which can be 
 seen in  lighter colors within the 
  total error bands.}  
\label{fig:a012}
\end{figure}

\begin{figure}[tbh]
\includegraphics[height=4.29cm,width=5.6cm]{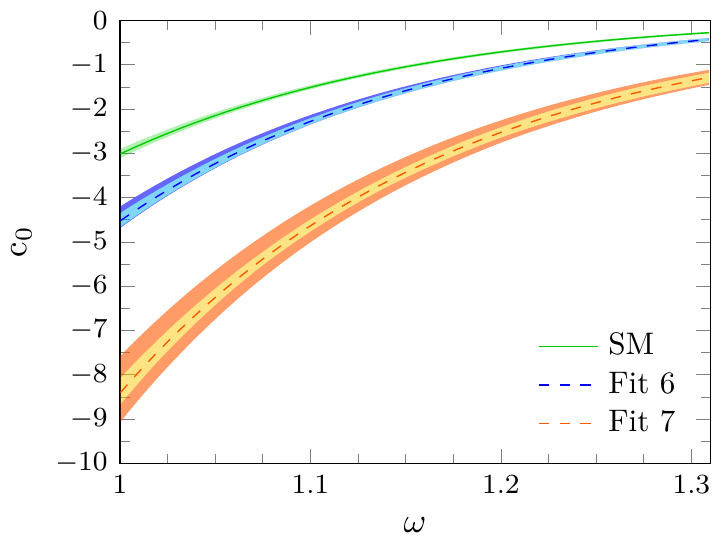}\hspace{.15cm} 
\includegraphics[height=4.15cm,width=5.6cm]{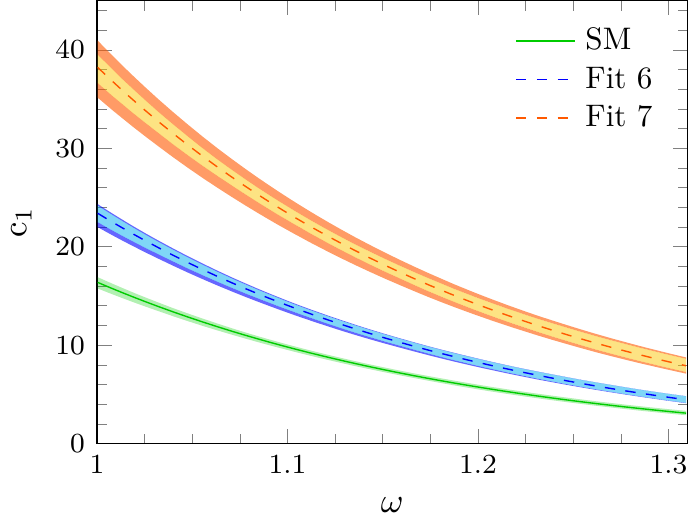} \hspace{.15cm}
\includegraphics[height=4.29cm,width=5.6cm]{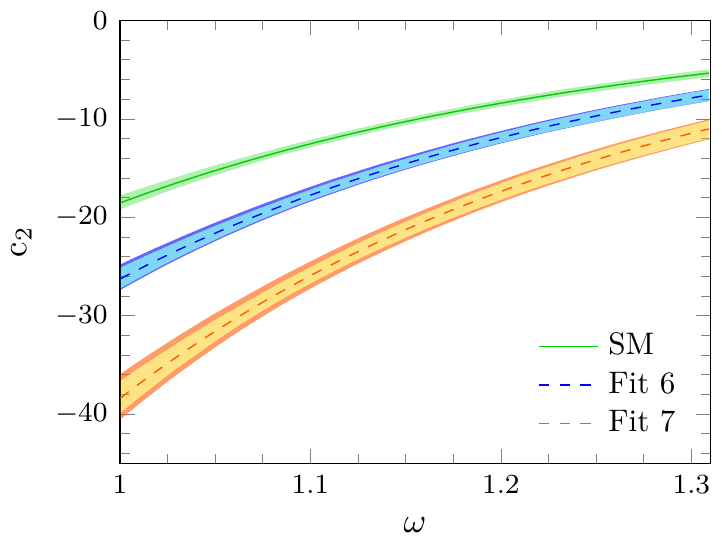}
\caption{LAB charged lepton energy expansion coefficients $c_0,c_1$ and $c_2$  (Eq.~\eqref{eq:ces}) for the 
unpolarized $d^2\Gamma[\Lambda_b\to\Lambda_c \tau\bar \nu_\tau]/(d\omega dE_\ell)$ differential decay width. Details 
as in  Fig.~\ref{fig:a012}.}  
\label{fig:c012}
\end{figure}

\begin{figure}[tbh]
\includegraphics[height=6cm]{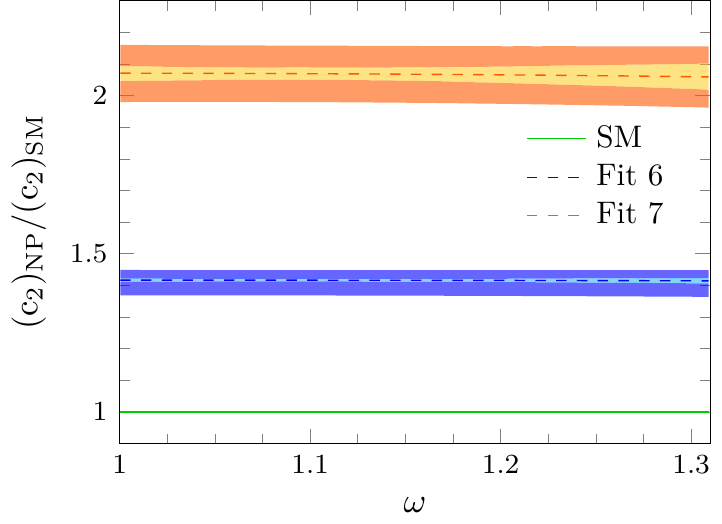}\hspace{.3cm}
\includegraphics[height=6cm]{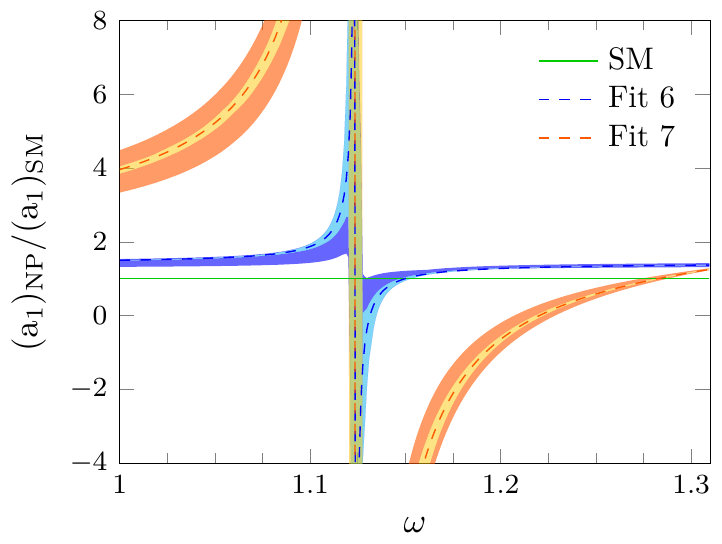}\\
\includegraphics[height=6cm]{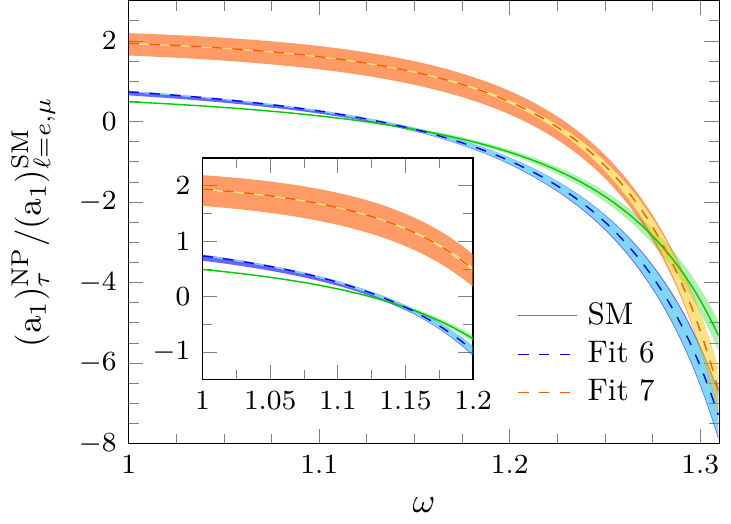}\hspace{.15cm}
\includegraphics[height=6cm]{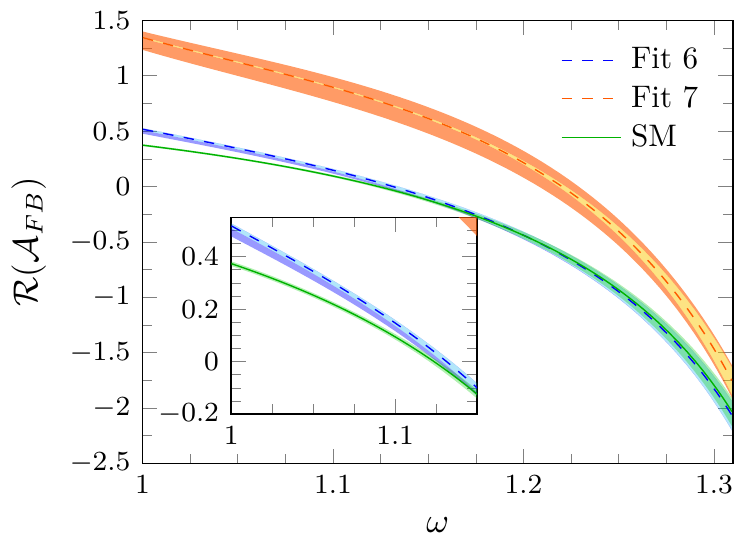}
\caption{ Top: 
$(c_2)_{\rm NP}/(c_2)_{\rm SM}$ (left) and $(a_1)_{\rm NP}/(a_1)_{\rm SM}$ (right) ratios for $\Lambda_b\to\Lambda_c \tau\bar \nu_\tau$, 
as a function of $\omega$.
Bottom: ${(a_1)_\tau^{\rm NP}(\omega)}/
 { (a_1)_{\ell=e,\mu}^{\rm SM}(\omega)}$  and ${\cal R}({\cal A}_{FB})$ ratios defined in Eqs.~\eqref{eq:ratioa1tauemu} and \eqref{eq:ratiosFB}, respectively. In the inserts to these latter plots, we
 amplify the $\omega$ region close to zero recoil. In all cases, we show results for the Fits 6 and 7 of Ref.~\cite{Murgui:2019czp}, and 
 details of the uncertainties are as in Fig.~\ref{fig:a012}.}  
\label{fig:c2ya1ratio}
\end{figure}

Fortunately, things improve considerably when one looks at the observables related
to the  CM $d^2\Gamma/(d\omega d\cos\theta_\ell)$ and LAB 
unpolarized $d^2\Gamma/(d\omega dE_\ell)$ double differential distributions. In Figs.~\ref{fig:a012} and
\ref{fig:c012} we show, respectively, the results for the $a_{i=0,1,2}$ and $c_{i=0,1,2}$ dimensionless
 coefficients that determine those distributions (Eqs.~\eqref{eq:as} and \eqref{eq:ces}). With the exception of 
$a_0$, the rest of these functions  allow a clear distinction between NP Fits 6 and 7 that
otherwise predict the same $d\Gamma/d\omega$ differential and total decay 
widths. We also display predictions, in the bottom panel of Fig.~\ref{fig:a012}, for the commonly used 
forward-backward asymmetry ${\cal A}_{FB}$, which features and $\omega-$behaviour are strongly determined by $a_1$.   If LFUV were experimentally established for the 
$\Lambda_b\to\Lambda_c$ semileptonic decays, the analysis of these observables 
would clearly help in establishing what kind of NP was  needed to reproduce 
experimental data.
Another way of presenting the results in Figs.~\ref{fig:a012} and 
\ref{fig:c012} is by showing  the ratios of the quantities obtained including NP
over their SM values. This is done for $a_1$ and  $c_2$ in 
Fig.~\ref{fig:c2ya1ratio}. We observe that the  $(c_2)_{\rm NP}/(c_2)_{\rm SM}$ 
ratio, depicted in the left-top panel of Fig.~\ref{fig:c2ya1ratio},
is almost constant having a very mild $\omega$ dependence.  However, it clearly 
distinguishes NP Fit 6 from Fit 7 and the two of them from the SM value of 1. We reached the same conclusions in our previous analysis of Ref.~\cite{Penalva:2019rgt}, but as it was the case with $d\Gamma/d\omega$, the proper consideration of the Wilson's coefficient statistical 
correlations  drastically reduces the errors in the predictions for this ratio\footnote{In fact, the reduction of uncertainties in this work 
compared to those given in \cite{Penalva:2019rgt} is very significant for all functions depicted in Figs.~\ref{fig:a012} and \ref{fig:c012}. }, which sharpens the NP discriminating power of this observable. 
Similar results  would be obtained for other ratios with the one for 
$a_1$, seen in the right-top panel of Fig.~\ref{fig:c2ya1ratio}, showing the stronger
 $\omega$ dependence. As seen in Figs.~\ref{fig:a012} and \ref{fig:c012}, 
 $a_1 (\omega)$ is the only function, of those shown in these two figures, 
 that presents a change of sign for the SM and the two NP scenarios analyzed in this work. 
This behaviour of $a_1$ explains the singularities in the 
NP ratios since,  for each model, the zeros of $a_1$ occur at different
 positions within the physical interval $[1, \omega_{\rm max}]$. 
 This strong $\omega-$dependence of the $(a_1)_{\rm NP}/(a_1)_{\rm SM}$ 
 ratio provides an additional NP-testing tool, which could be used when 
 future accurate measurements are available.

 An alternative to this latter ratio that can be obtained just from pure experimental data
 is the following. Assuming that NP affects only the third generation 
 of leptons,  $a_1$ for  $\ell=e,\mu$ (that can be considered as 
 massless to a high degree of approximation)
 is a pure SM result,  and the 
 ratio $(a_1)_\tau^{\rm NP}/(a_1)_{\ell=e,\mu}^{\rm SM}$ can be measured 
 from the asymmetry between the number of events observed for $\theta_\ell \in [0,\pi/2]$ and  for  $\theta_\ell \in [\pi/2, \pi]$,  
\begin{equation}
 \frac{(a_1)_\tau^{\rm NP}(\omega) \tstrut}
 {\tstrut (a_1)_{\ell=e,\mu}^{\rm SM}(\omega) } = 
 \left(1-\frac{m_\tau^2}{q^2}\right)^{-2}\  \frac{\bigintss_0^1 d\cos\theta_{\ell}\left[\frac{d^2\Gamma[\Lambda_b\to\Lambda_c \tau\bar \nu_\tau]}{d\omega d\cos\theta_\ell}(\omega, \theta_\ell)\right]-\bigintss_{-1}^0 d\cos\theta_{\ell} \left[\frac{d^2\Gamma[\Lambda_b\to\Lambda_c \tau\bar \nu_\tau]}{d\omega d\cos\theta_\ell}(\omega, \theta_\ell)\right]\tstrut}{\tstrut \bigintss_0^1 d\cos\theta_{\ell}\left[\frac{d^2\Gamma[\Lambda_b\to\Lambda_c e(\mu)\bar \nu_{e(\mu)}]}{d\omega d\cos\theta_\ell}(\omega, \theta_\ell)\right]-\bigintss_{-1}^0 d\cos\theta_{\ell}\left[\frac{d^2\Gamma[\Lambda_b\to\Lambda_c e(\mu)\bar \nu_{e(\mu)}]}{d\omega d\cos\theta_\ell}(\omega, \theta_\ell)\right]} 
 \label{eq:ratioa1tauemu}
\end{equation}
This ratio is shown in the left-bottom panel of Fig.~\ref{fig:c2ya1ratio}. Since 
$(a_1)_{\ell=e,\mu}^{\rm SM}(\omega)$ does not vanish for $\omega>1$ (see Fig.~1 of Ref.~\cite{Penalva:2019rgt}),
no divergence appears in this case. In the insert to this latter panel we amplify
the $\omega\in[1,1.2]$ region to better show the discriminating power of this observable
close to zero recoil.
To minimize experimental and theoretical uncertainties, both the numerator and 
the denominator of the right hand side of Eq.~\eqref{eq:ratioa1tauemu} can be normalized by $d\Gamma/d\omega$ for each decay mode. 
In this way, the ratio ${\cal R}({\cal A_{FB}})$, defined as
\begin{equation}
{\cal R}({\cal A}_{FB}) = \frac{({\cal A}_{FB})_\tau^{\rm NP}}{({\cal A}_{FB})_{\ell=e,\mu}^{\rm SM}\tstrut}= 
\frac{ \Big[ \frac{a_1 \tstrut}{2 a_0 +2a_2/3} \Big]_\tau^{\rm NP}\tstrut}
 {\tstrut \Big[ \frac{a_1 \tstrut}{2 a_0 +2a_2/3} \Big]_{\ell=e,\mu}^{\rm SM} } \label{eq:ratiosFB}
\end{equation}
can be measured by subtracting the number of events seen for $\theta_\ell \in [0,\pi/2]$ and  
for  $\theta_\ell \in [\pi/2, \pi]$ and dividing by the total sum of observed events, in each of 
the $\Lambda_b\to \Lambda_c \tau\bar \nu_\tau$ and $\Lambda_b\to \Lambda_c e(\mu)\bar \nu_{e(\mu)}$ reactions. We expect this strategy should remove 
a good part of experimental normalization errors. We show the theoretical predictions for 
${\cal R}({\cal A}_{FB})$ in the bottom-right panel of Fig.~\ref{fig:c2ya1ratio}, where we see a significant reduction of uncertainties, and 
the potential of this ratio to establish the validity of the NP scenarios 
associated to Fit 7.  To avoid confusion, we must warn the reader that 
${\cal R}({\cal A_{FB}})$ introduced here is not related  with a  ratio 
of hadronic forward-backward asymmetries defined in Eq.~(2.46) of 
Ref.~\cite{Boer:2019zmp}, and which is discussed in Fig.1 of that work. 
The angles used in  \cite{Boer:2019zmp} are different to those employed 
in the present analysis.

\begin{figure}[tbh]
\includegraphics[height=4cm,width=5.6cm]{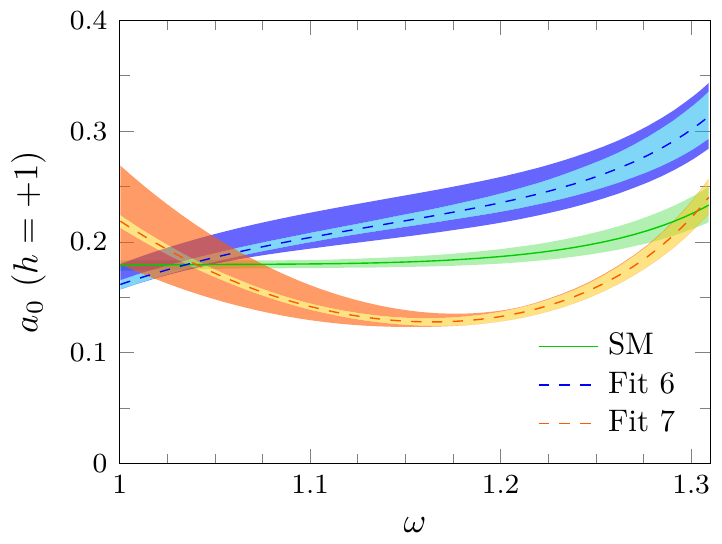}\hspace{.25cm}
\includegraphics[height=4cm,width=5.6cm]{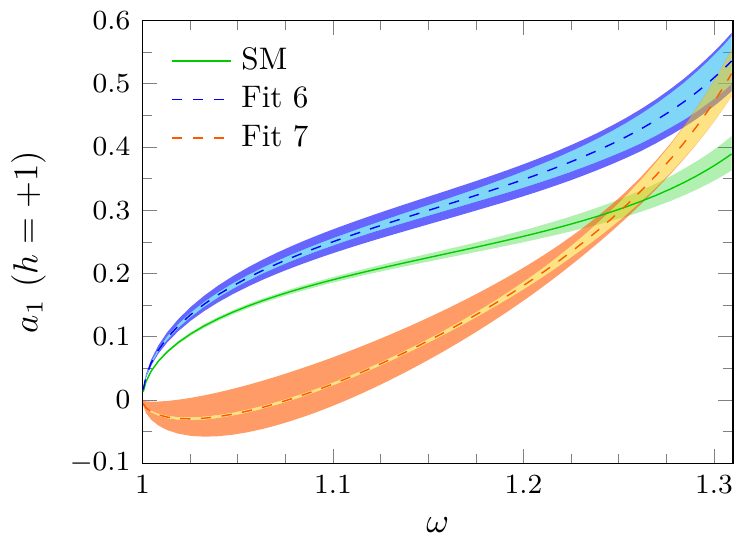}\hspace{.25cm}
\includegraphics[height=4cm,width=5.6cm]{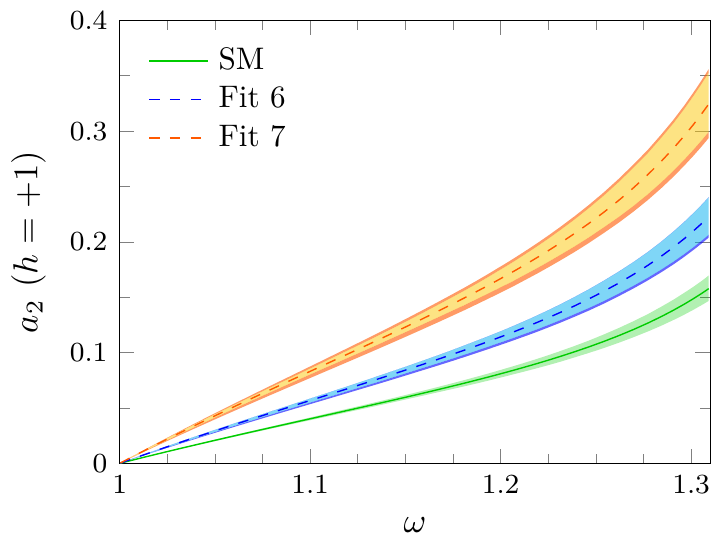}\\
\includegraphics[height=4cm,width=5.6cm]{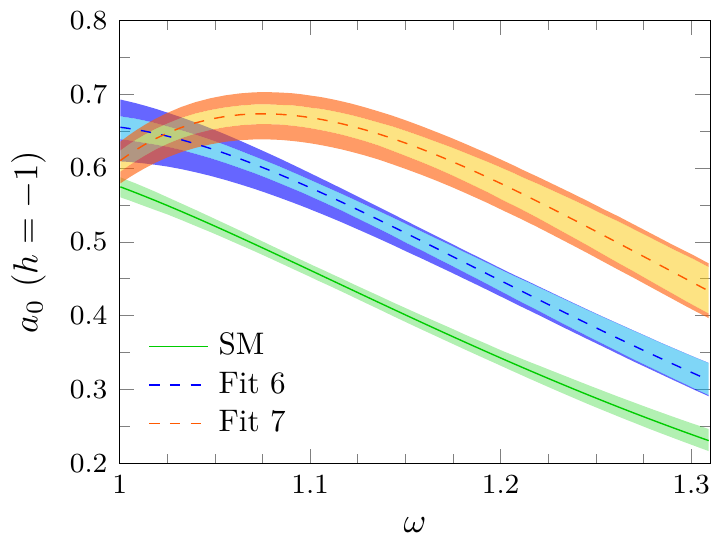}\hspace{.25cm}
\includegraphics[height=4cm,width=5.6cm]{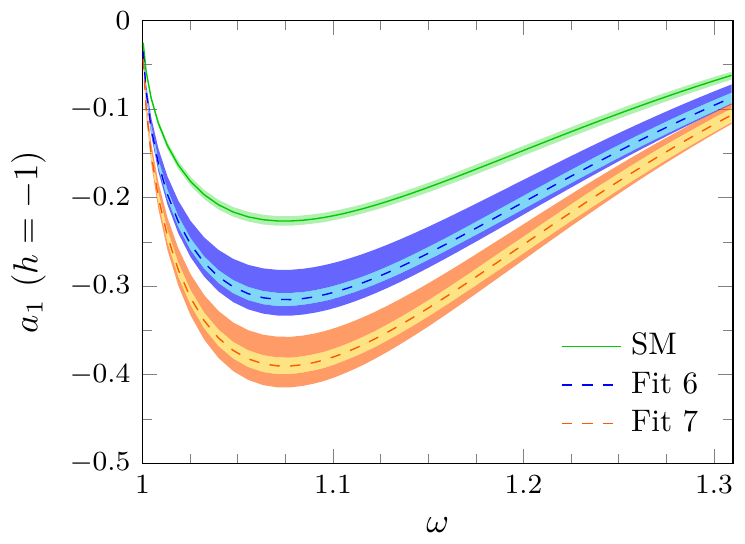}\hspace{.25cm}
\includegraphics[height=4cm,width=5.6cm]{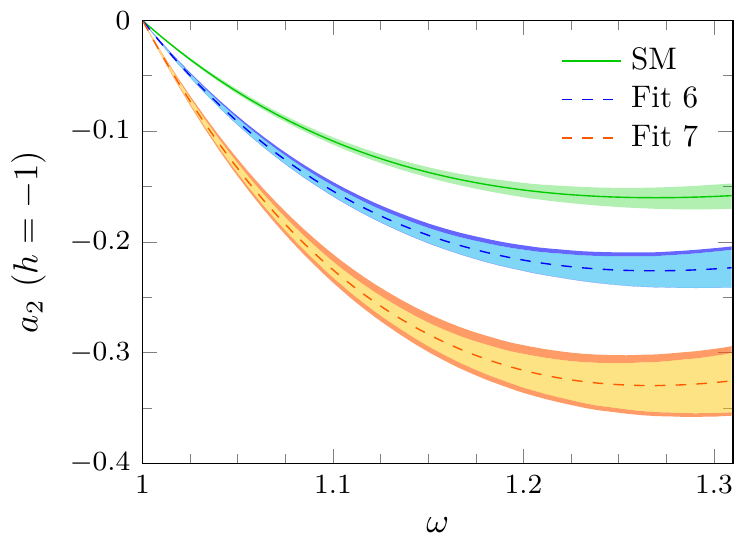}\\
\caption{ CM angular coefficients for positive and negative helicities $\left(a_{i=0,1,2}(h=\pm 1)\right)$ for the $\tau-$mode $\Lambda_b\to \Lambda_c$ semileptonic decay, as a function of $\omega$. Details as in Fig.~\ref{fig:a012}.}  
\label{fig:a012pol}
\end{figure}

\begin{figure}[tbh]
\includegraphics[height=4cm,width=5.6cm]{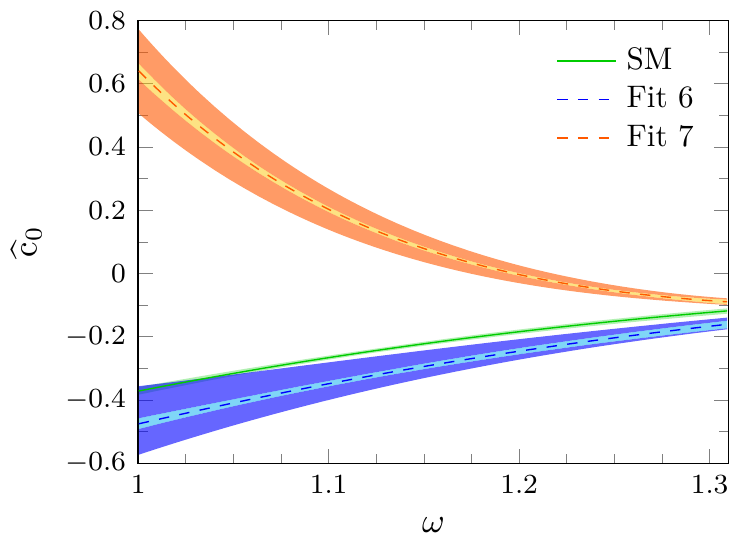}\hspace{.3cm}\includegraphics[height=4cm,width=5.6cm]{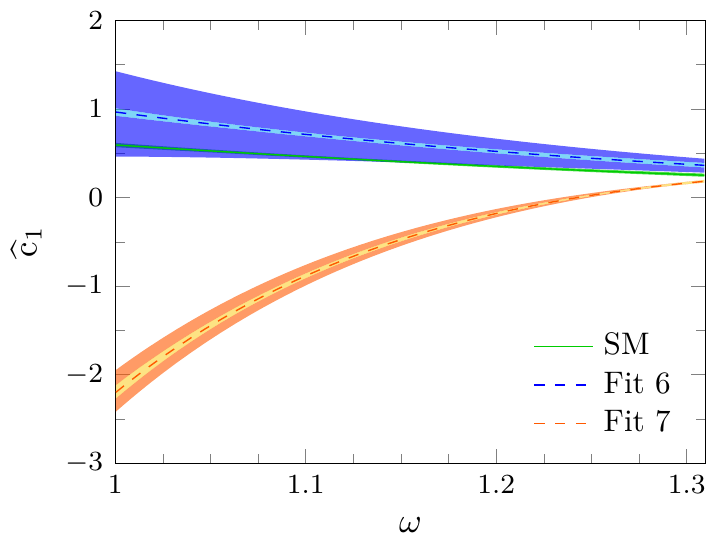}\hspace{.3cm}\includegraphics[height=4cm,width=5.6cm]{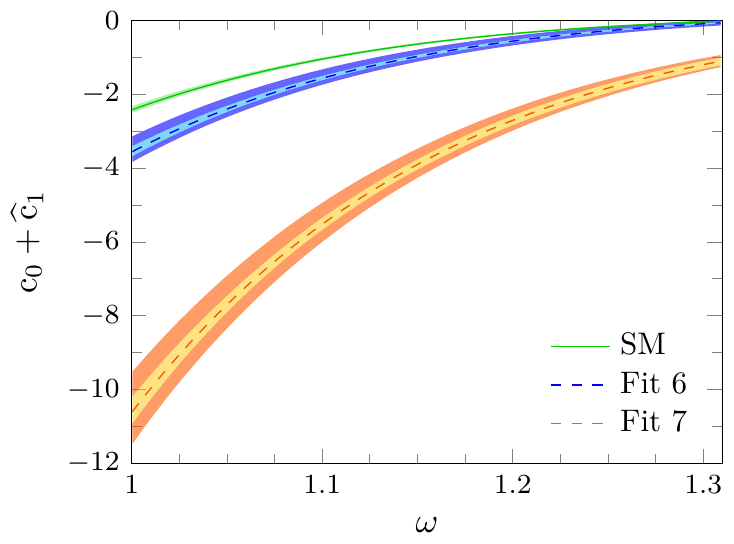} \\\vspace{0.25cm}
\includegraphics[height=4cm,width=4.2cm]{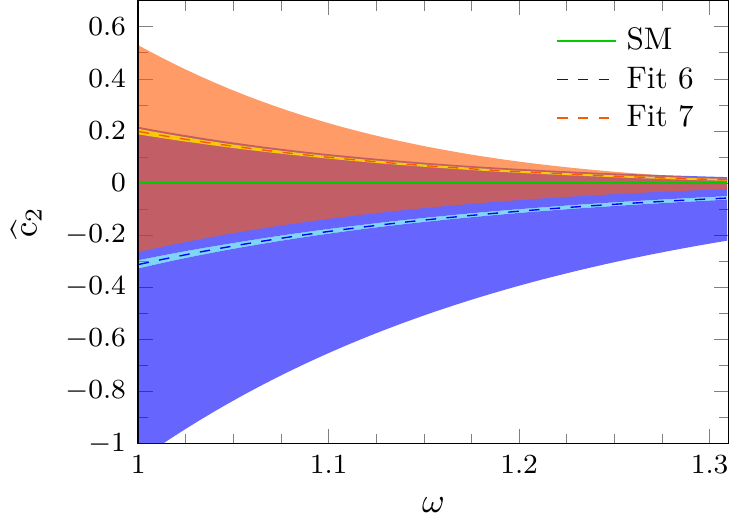}\hspace{.2cm}\includegraphics[height=4cm,width=4.2cm]{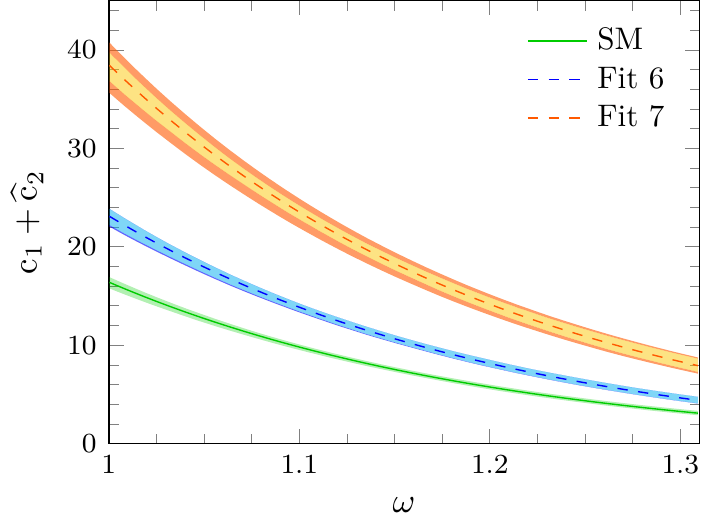} \hspace{0.2cm}\includegraphics[height=4.2cm,width=4.2cm]{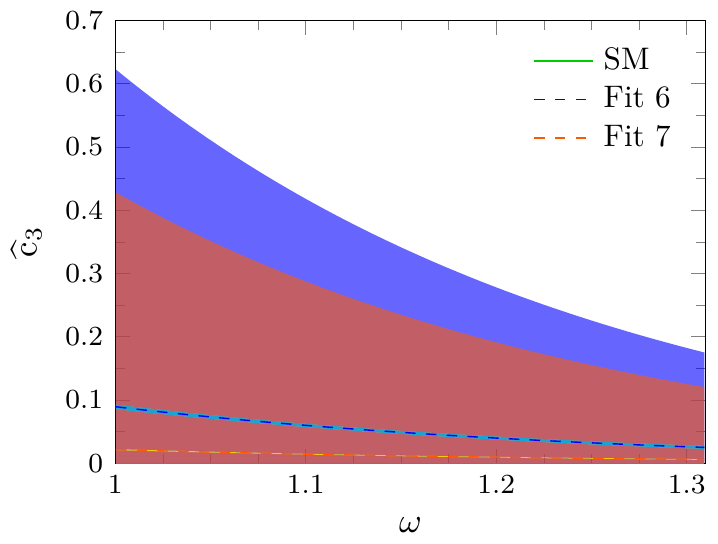}\hspace{0.2cm}\includegraphics[height=4.2cm,width=4.2cm]{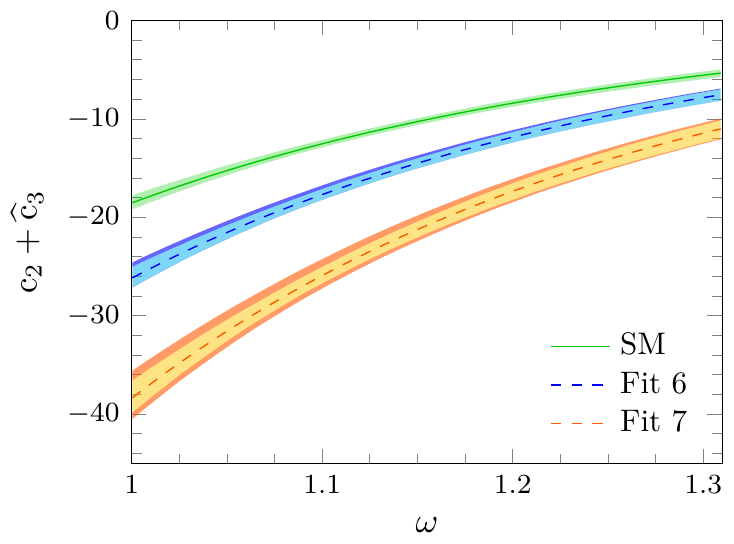}
\caption{LAB charged lepton energy expansion coefficients $\widehat c_{i=0,1,2,3}$ (Eqs.~\eqref{eq:Chpol} and \eqref{eq:cespol}) for the 
polarized $d^2\Gamma[\Lambda_b\to\Lambda_c \tau\bar \nu_\tau]/(d\omega dE_\ell)$ differential decay width. We also show the $(c_0+\widehat c_1)$, $(c_1+\widehat c_2)$ and $(c_2+\widehat c_3)$ sums in the third top, second and fourth bottom panels, respectively. Details  as in Fig.~\ref{fig:a012}.}  
\label{fig:c0123hat}
\end{figure}

To complete the analysis, we display in Figs.~\ref{fig:a012pol} and \ref{fig:c0123hat} additional predictions for the 
polarized CM $d^2\Gamma/(d\omega d\cos\theta_\ell)$ and LAB $d^2\Gamma/(d\omega dE_\ell)$ distributions. As in the previous figures, we separate in all the observables the errors produced by the uncertainties in the LQCD determination of the form factors, which for the NP results are not negligible at all, and become even dominant in certain cases. In Fig.~\ref{fig:a012pol},  we show  
the CM angular coefficients for positive and negative helicities, $a_{i=0,1,2}(h=\pm 1)$, which explicit expressions in terms of the $\widetilde W$ SFs were compiled in  Eq.~\eqref{eq:aspol}. Even taking uncertainties into account, Fits 6 and 7 provide
distinctive predictions that also differ from the SM results. We see that $(h=+1)$ and  $(h=-1)$ coefficients are comparable in size, and we systematically find  
\begin{equation}
 \left |a_{0,1,2}^{\rm NP-Fit\ 7}(h=-1)\right| \ge \left|a_{0,1,2}^{\rm NP-Fit\ 6}(h=-1)\right|\ge \left|a_{0,1,2}^{\rm SM}(h=-1)\right|
\end{equation}
except for $a_0(h=-1)$ in a narrow region, $\omega=1-1.03$, where the NP Fit 6 and 7 predictions agree within errors. In the case of 
$a_0(h=+1)$ and $a_1(h=+1)$, roughly,  NP Fit 6 values are greater than the Fit 7 ones, with SM results in the middle. Note that 
the partial integrated rates, $d\Gamma/d\omega$ shown in Fig.~\ref{fig:dgdw1}, are not sensitive to the $a_1-$contributions, and therefore having access to the detailed angular dependence provides very valuable additional information. 
We also see large cancellations in  $a_2=a_2(h=+1)+a_2(h=-1)$, which become total, both at zero recoil and at the 
end of the phase space, where the sum $a_2$ vanishes. Actually for $\omega=\omega_{\rm max}$, $|a_2(h=\pm 1)|$ are as big as
$a_0(h=\pm 1)$.
In Fig.~\ref{fig:c0123hat} we show the  $\widehat c_{i=0,1,2,3}$ coefficients, given in Eq.~\eqref{eq:cespol}, that appear in the expression for the 
polarized LAB $d^2\Gamma/(d\omega dE_\ell)$ double differential decay width. Taking into account the uncertainty bands, dominated by the errors of the Wilson coefficients, only $\widehat c_0$ and $\widehat c_1$ can be used to distinguish between NP Fits 6 and 7, while only Fit 7 predicts a result in clear disagreement with SM expectations. We see NP Fits 6 and 7 predictions for these  two coefficients have even opposite signs, and the differences are enhanced in the sum $(c_0+\widehat c_1)$, which is the coefficient of the linear $E_\ell$ term in Eq.~\eqref{eq:Chpol}. 

The other two observables $\widehat c_2 $ and $\widehat c_3$ are of little use for the current analysis, because the results of Fits 6 and 
7 overlap and, furthermore,  these coefficients are around two orders of magnitude lower than $c_1 $ and $ c_2$, respectively. One should 
note that $\widehat c_2 $ and $\widehat c_3$ are proportional to the tensor-diagonal $\widetilde W^T_{2,4}$ and tensor-interference $\widetilde W_{I_3,I_5,I_6}$ SFs, and therefore both are zero in the SM. Moreover, for the NP scenarios associated to Fit 6 and 7 of Ref.~\cite{Murgui:2019czp}, these two coefficients of the unpolarized distribution are negligible, since for both fits $|C_T|$ is already of the order of $10^{-2}$, and compatible with zero, $C_T=0.01^{+0.09}_{-0.07}$ and $-0.02^{+0.08}_{-0.07}$, respectively. 

However, it is important to stress that $\widehat c_2 $ and $\widehat c_3$ are optimal observables to restrict the validity of  NP schemes with high tensor contributions. As a matter of example, 
\begin{equation}
 \frac{\widehat{c}_3(\omega)}{c_2(\omega)}= \frac{32\widetilde W_2^T } 
 {\widetilde W_2-\tstrut16 \widetilde W_2^T }= 
 \frac{32\widetilde W_2^T}{\tstrut\widetilde W_2}+\cdots = -32 x  +{\cal O}(\Lambda_{QCD}/m_{b,c}), \quad 
 x= \frac{2 |C_T|^2}{|C_V|^2+ |C_A|^2}
\end{equation}
where we have made used that $\widetilde W_2^T/\widetilde W_2 = -x+{\cal O}(\Lambda_{QCD}/m_{b,c})$, as deduced from Eq.~\eqref{eq:hqss}.

On the other hand, the small NP tensor contribution for Fits 6 and 7, together with the heavy quark limit relations of Eq.~\eqref{eq:hqss}, explains the flat $\omega-$behavior of the $(c_2)_{NP}/(c_2)_{SM}$ ratio seen in Fig.~\ref{fig:c2ya1ratio}. If one neglects $\widetilde W^T_{2}$, the coefficient $c_2$ is proportional   to $\widetilde W_2$. The linear  $C_{V_R}$ terms, that could induce a non-zero $\omega$ 
dependence in $(\widetilde W_2)_{NP}/(\widetilde W_2)_{SM}$, cancel to 
order ${\cal O}(\Lambda_{\rm QCD}/m_{b,c})$. 

\begin{figure}[tbh]
\includegraphics[height=4.5cm]{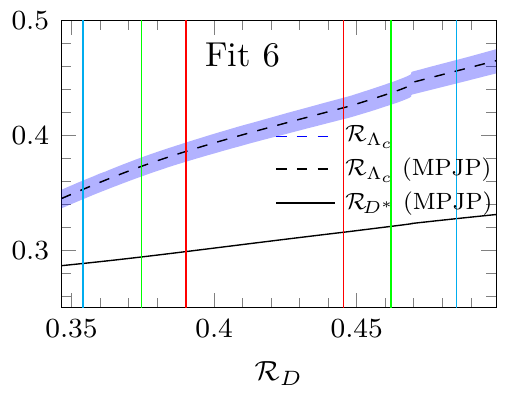}\hspace{.15cm} 
\includegraphics[height=4.5cm]{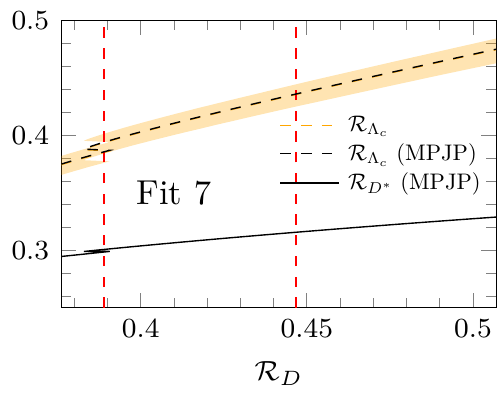} \hspace{.15cm} 
\includegraphics[height=4.5cm]{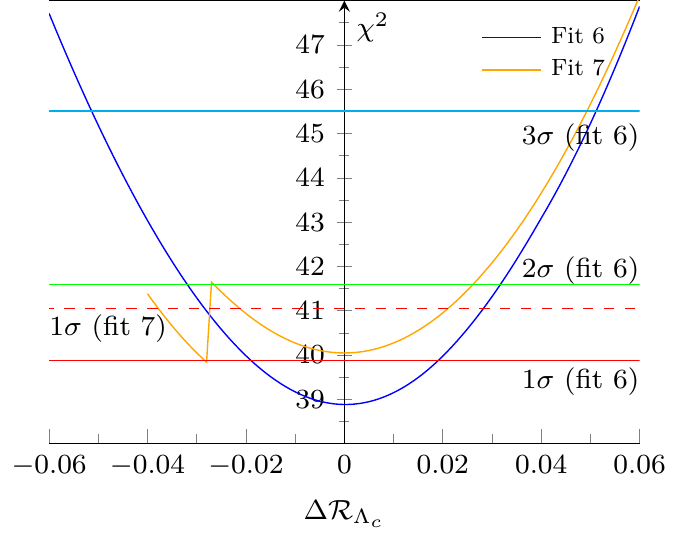}
\caption{ Left/middle panel: NP Fits 6 and  7 results for ${\cal R}_{\Lambda_c}$ and ${\cal R}_{D^{*}}$ 
as a function of ${\cal R}_D$ for different $\chi^2-$weighted samples of Wilson coefficients.  Black solid and dashed curves, 
labeled as MPJP, stand for the results of Ref.~\cite{Murgui:2019czp} provided by the authors of that work~\cite{Ana}. The  blue and orange dashed 
lines (indistinguishable from the MPJP predictions) correspond to the current numerical evaluation of ${\cal R}_{\Lambda_c}$ for Fits 6 and 7, respectively, with the shaded bands
showing the 68\% CL uncertainties inherited from the LQCD determination of the form-factors~\cite{Detmold:2015aaa, Datta:2017aue}. Right panel: Chi-square values~\cite{Murgui:2019czp,Ana} for each set of Wilson coefficients (sWC) used in the left and middle panels, and represented in this
  plot by $\Delta{\cal R}_{\Lambda_c}={\cal R}_{\Lambda_c}^{\rm sWC}-{\cal R}_{\Lambda_c}^{\min}$, with ${\cal R}_{\Lambda_c}^{\min}= 0.405$ 
  and 0.415 for Fits 6 and 7, respectively.}  
\label{fig:rl}
\end{figure}

Finally, in Fig.~\ref{fig:rl}, we present ${\cal R}_{\Lambda_c}$ as a function of ${\cal R}_D$ obtained using NP Fit 6 (left) and  7 (middle) $\chi^2-$weighted samples  of Wilson coefficients provided by the authors of Ref.~\cite{Murgui:2019czp}. In  Fig.~\ref{fig:rl},  we include sets beyond the $1\sigma$ ones. For illustration purposes, we also show  the results of Ref.~\cite{Murgui:2019czp} for the ratio ${\cal R}_{D^*}$, which allows us to highlight the clear correlation between these three LFUV observables. Note that 
SM predictions for the ${\cal R}_D^{\rm SM}=0.300 \pm 0.05$ and ${\cal R}_{D^*}^{\rm SM}=0.251 \pm 0.004$ ratios are below the ranges considered, while ${\cal R}^{\rm SM}_{\Lambda_c} = 0.332 \pm 0.008$.
In the right panel  of  Fig.~\ref{fig:rl}, we show, for each of the  Wilson coefficient sets  used in the left and middle panels, the  $\chi^2-$variations   against the corresponding changes induced in the ${\cal R}_{\Lambda_c}$ ratio\footnote{There exist one-to-one relations between
each set of Wilson coefficients (sWC) used in the left (Fit 6) and
middle (Fit 7) panels of Fig.~\ref{fig:rl} and the chi-square
values or the variations $\Delta{\cal R}_{\Lambda_c}(={\cal
  R}_{\Lambda_c}^{\rm sWC}-{\cal R}_{\Lambda_c}^{\min})$ shown in the
right plot of the figure. At some point for $\Delta {\cal
  R}_{\Lambda_c} < -0.02$, the local Fit 7 collapses into Fit 6.}. Both, Fit 6 and Fit 7 $\chi^2$ functions grow from their 
minimum values, and the $\Delta\chi^2=1$, $\Delta\chi^2=2.71$, $\Delta\chi^2=6.63, \cdots$ increments can be 
used to determine the  68\% ($1\sigma$),  90\% ($2\sigma$), 99\%($3\sigma$), $\cdots$  CL intervals of the NP predictions for ${\cal R}_{D}$, 
${\cal R}_{D^*}$ and ${\cal R}_{\Lambda_c}$. 

\section{Conclusions}
\label{sec:conclusions}
We have included the NP tensor term, and all
the interference contributions associated with it, in our general formalism 
for semileptonic decays initially introduced in Ref.~\cite{Penalva:2019rgt}.  In this way,  all the NP effective Hamiltonians that are 
considered in LFUV studies with massless left-handed neutrinos have now been taken into account, including the possibility of violation of CP-symmetry due to the presence of
complex Wilson coefficients. The scheme developed is totally general and it can be
 applied to any charged current semileptonic decay, involving any quark 
 flavors or initial and final hadron states. 
 
We have shown that a total of sixteen  SFs ($\widetilde W$'s) are needed to fully describe the hadronic tensor.
They are constructed out of the complex Wilson coefficients, that characterize the strength of
the different NP terms, and the form factors needed to describe the genuine hadronic 
matrix elements.  We have also derived  general expressions for unpolarized and charged-lepton polarized 
CM $d^2\Gamma/(d\omega d\cos\theta_\ell)$ and LAB $d^2\Gamma/(d\omega dE_\ell)$ differential decay widths in terms of the $\widetilde W-$SFs. 
Unlike the unpolarized case, where all the accessible observables could be determined either from the CM 
 or LAB distributions, we have pointed out  that LAB and CM charged lepton helicity distributions should be used  simultaneously, since in the polarized case,  they provide complementary  information.  We have also shown that, even assuming that the NP terms affect
 only to the third lepton family, the strategy  to obtain full polarized information for tau-mode decays from non-polarized $e/\mu$ and $\tau$ 
 data is spoiled by the presence of NP tensor operators.  

As a result of this general discussion, we have concluded that determining all NP  parameters, with their complex phases,  
from  a single type of decay is tough,  even assuming  that the hadronic form factors are known. This is because the  experimental measurements 
of the required polarized decays are very challenging due to the presence of undetected neutrinos.  We have argued it is therefore essential  to simultaneously analyze data from various 
types of semileptonic decays  (f.e. $\bar B\to D, \bar B\to D^*$,
 $\Lambda_b\to \Lambda_c, \bar B_c\to \eta_c, \bar B_c\to J/\Psi$...), 
 considering both the $e/\mu$ and $\tau$ modes, and that the scheme presented in this work is a powerful tool to achieve this goal.
 
 The general formalism developed  has then been applied
to update the analysis of the $\Lambda_b\to\Lambda_c\tau\bar\nu_\tau$ decay carried out in Ref.~\cite{Penalva:2019rgt}. 
We have found small numerical differences for central results, because of the little strength of the tensor terms in the  
NP scenarios originally  considered in our previous work. However, the proper consideration of the Wilson's coefficient statistical 
correlations  has drastically reduced the errors in the new predictions, which have significantly improved 
the NP discriminating power of the present study. In addition, we have obtained full results for both CM and LAB charged lepton polarized distributions.  

As in Ref.~\cite{Penalva:2019rgt}, we have shown the potential of the 
CM $d^2\Gamma/(d\omega d\cos\theta_\ell)$ and LAB  $d^2\Gamma/(d\omega dE_\ell)$
 distributions to distinguish between 
 models, fitted to $b\to c\tau\bar\nu_\tau$ anomalies in the meson sector,  
 that differ in the  strengths of the NP terms but that otherwise give the
 same differential $d\Gamma/d\omega$ and integrated decay widths. In particular,
  the  $a_1$ and $a_2$, and   all three $c_0,\,c_1$ and $c_2$ functions, 
  associated with the  non-polarized  
  CM $d^2\Gamma/(d\omega d\cos\theta_\ell)$ and LAB $d^2\Gamma/(d\omega dE_\ell)$ distributions, respectively, 
  are very well suited for that purpose, 
at least for the $\Lambda_b\to \Lambda_c$ semileptonic decay specifically 
studied in this work. For this baryon transition, we have also shown the great interest of 
the ratios $(a_1)_\tau^{\rm NP}/(a_1)_{\ell=e,\mu}^{\rm SM}$ and 
${\cal R}({\cal A}_{FB})$ (Eqs.~\eqref{eq:ratioa1tauemu} and \eqref{eq:ratiosFB}, respectively) which can be directly measured  from the $(\theta_\ell, \pi-\theta_\ell)$ asymmetry of the experimental distributions.   
If LFUV  is experimentally established for this decay, the 
analysis of all these observables 
can help in understanding what kind of NP is needed to  explain the data. 
Finally, we have identified two coefficient functions, in the LAB polarized 
distribution, which theoretically should be very efficient in restricting 
the validity of NP schemes with a sizable tensor contribution, although we 
are aware of the  difficulty  of their measurement at present and in 
the near future.

\section*{Acknowledgements}
We warmly thank C. Murgui, A. Pe\~nuelas  and A. Pich for useful
discussions. N. P. and J.N. want to acknowledge the hospitality and financial support of the Nuclear
Physics Group of the University of Salamanca. This research has been supported  by the Spanish Ministerio de
Econom\'ia y Competitividad (MINECO) and the European Regional
Development Fund (ERDF) under contracts FIS2017-84038-C2-1-P, 
FPA2016-77177-C2-2-P, 
and by the EU Horizon 2020 
research and innovation programme, STRONG-2020 project, under grant agreement 
No 824093. 
\appendix
\section{Lepton tensors}
\label{sec:appL}

From Eq.~\eqref{eq:lepton-current}, in the limit of massless neutrinos, 
we obtain
\be
J^{L}_{(\alpha\beta)}(k,k';h)[J^{L}_{(\rho\lambda)}(k,k';h)]^* = 
\frac14 {\rm Tr}\Big[(\slashed{k}'+ m_{\ell}) 
\Gamma_{(\alpha\beta)} (1-\gamma_5)\slashed{k}\,\widetilde
\Gamma_{(\rho\lambda)}P_h\Big], \qquad \widetilde\Gamma_{(\rho\lambda)} = 
\gamma^0 \Gamma_{(\rho\lambda)}^\dagger \gamma^0.\label{eq:LT}
\ee
The different $\Gamma_{(\alpha\beta)}$ and $\Gamma_{(\rho\lambda)}$ operators give rise to the following lepton tensors (we use the convention $\epsilon_{0123}=+1$ and $g_{\mu\nu}=(+,-,-,-)$)
\begin{eqnarray}
L(k,k';h) &=& \left(k\cdot k' + h\,\, k\cdot s\right)/2, \label{eq:leptensorSP}\\
L_\alpha(k,k';h)&=&\frac{m_\ell}{2} k_\alpha + \frac{h}{2 m_\ell}
\left(k'_\alpha\, k\cdot s-s_\alpha\, k\cdot k' + i
\epsilon_{\alpha\delta\eta\sigma}k^{\prime \delta}k^\eta s^\sigma\right), \label{eq:leptensor-vec}\\
L'_{\rho\lambda}(k,k';h)&=& \frac{i}{2 } \left(k_{\rho} k'_{\lambda}-k_{\lambda}
k'_{\rho} +i\epsilon_{\rho\lambda\delta\eta}k'^{\delta} k^{\eta}\right) + i
\frac{h}2 \left( k_\rho s_\lambda- k_\lambda
s_\rho+i\epsilon_{\rho\lambda\delta\eta}s^\delta k^\eta \right),  \label{eq:leptensor-ST}\\
L_{\alpha\rho}(k,k';h) &=& \frac12  \left(k'_\alpha k_\rho+k_\alpha k'_\rho -
g_{\alpha\rho} k\cdot k'+ i\epsilon_{\alpha\rho\delta\eta}k'^{\delta}
k^{\eta}\right) - \frac{h}{2} \left ( s_{\alpha} k_{\rho}+k_{\alpha} s_{\rho} -
g_{\alpha\rho} k\cdot s+ i\epsilon_{\alpha\rho\delta\eta}s^{\delta}
k^{\eta}\right),\label{eq:leptensorVA}\\
L_{\alpha\rho\lambda}(k,k';h) &=& \frac{i m_{\ell}}{2}  \left( g_{\alpha\lambda}k_\rho
-g_{\alpha\rho}k_\lambda + i \epsilon_{\alpha\rho\lambda\delta}k^\delta\right) 
\nonumber\\&&- \frac{i h}{2 m_\ell} \Big[ k'_\alpha (s_\rho k_\lambda - s_\lambda k_\rho)
+ k_\alpha (s_\rho k'_\lambda - s_\lambda k'_\rho) +
 s_\alpha (k_\rho k'_\lambda - k_\lambda k'_\rho) \nonumber \\
 &&\hspace{1.5cm}+ (k\cdot k') 
(g_{\alpha\rho}s_\lambda -g_{\alpha\lambda}s_\rho ) + (s\cdot k) 
(g_{\alpha\lambda}k'_\rho -g_{\alpha\rho}k'_\lambda )\Big]
\nonumber\\
&&-\frac{h}{2m_\ell} \Big [ (k\cdot k') \epsilon_{\alpha \rho\lambda \delta } 
s^\delta  + s_\lambda \epsilon_{\alpha\rho\delta\eta}k^{\prime\delta}k^\eta 
- s_\rho \epsilon_{\alpha\lambda\delta\eta}k^{\prime\delta}k^\eta + k_\alpha 
\epsilon_{\rho\lambda\delta\eta}s^\delta k^{\prime \eta }+ k'_\alpha 
\epsilon_{\rho\lambda\delta\eta}s^\delta k^{\eta } \Big], \label{eq:lepdif}
\\
L_{\alpha\beta\rho\lambda}(k,k';h) & =& \frac12 L_{\alpha\beta\rho\lambda}(k,k')
+ \frac{h}{2}L_{\alpha\beta\rho\lambda}(k,s), \label{eq:leptensor}
\end{eqnarray}
which correspond to the $(\Gamma_{(\alpha\beta)}, \Gamma_{(\rho\lambda)})=(1, 1), (\gamma_\alpha,1), (1,\sigma_{\rho\lambda}), (\gamma_\alpha,\gamma_\rho)$, $(\gamma_\alpha,\sigma_{\rho\lambda})$, and $(\sigma_{\alpha\beta},\sigma_{\rho\lambda})$ combinations, respectively, and in Eq.~\eqref{eq:leptensor}
\begin{eqnarray}
 L_{\alpha\beta\rho\lambda}(k,k') &=& g_{\beta \rho} (k_\alpha k'_\lambda+ k_\lambda k'_\alpha)- g_{\beta \lambda} (k_\alpha k'_\rho+ k_\rho k'_\alpha) 
 -g_{\alpha \rho} (k_\beta k'_\lambda+ k_\lambda k'_\beta)+ g_{\alpha \lambda} (k_\beta k'_\rho+ k_\rho k'_\beta) \nonumber \\
 &&+ (k\cdot k')(g_{\alpha \rho}g_{\beta \lambda} 
 -g_{\alpha \lambda}g_{\beta \rho}) +i 
 \Big(k'_\alpha \epsilon_{\beta\lambda\rho\delta} k^\delta- 
 k'_\beta \epsilon_{\alpha\lambda\rho\delta} k^\delta + k_\rho 
 \epsilon_{\alpha\beta\lambda\delta} k^{\prime \delta}- k_\lambda 
 \epsilon_{\alpha\beta\rho\delta} k^{\prime \delta} \Big).
\end{eqnarray}

\newpage

\section{Hadron tensors}
\label{sec:appH}

We collect here the hadron tensors that should be contracted with  the
corresponding lepton ones, compiled in the previous appendix,  to obtain 
$\overline\sum  |{\cal M}|^2$. 
In Sec.~\ref{sec:hme}, we have addressed the diagonal 
$J_{H}^\alpha [J_{H}^\rho]^*$ case. In this appendix, we begin with the 
diagonal $J_{H}^{\alpha\beta} [J_{H}^{\rho\lambda}]^*$ tensor term, which it 
is also discussed in detail. The decomposition of the rest of the hadron tensors
 as linear combination of independent Lorentz (pseudo-)tensor 
 structures\footnote{They are constructed out the vectors $p^\mu$, $q^\mu$, 
 the metric $g^{\mu\nu}$ and the Levi-Civita pseudotensor 
 $\epsilon^{\mu\nu\delta\eta}$.} is listed after, and it is obtained similarly
  to those in the two previous examples. The coefficients multiplying   
  the (pseudo-)tensors are the  $\widetilde W's$ SFs, which depend on $q^2$ 
  and the hadron masses.  As mentioned in the Introduction, there appear 16 
  $\widetilde W's$ SFs, which are
 constructed out of  NP complex Wilson coefficients  and the  genuine 
 hadronic responses ($W's$). The latter ones are determined 
 by the matrix elements of the involved hadron operators, which for each 
 particular decay are parametrized in terms of form-factors.

\begin{itemize}
 \item The diagonal contribution of the tensor operator $O_H^{\alpha\beta}$ 
 gives rise to a (pseudo-)tensor of four indices
\be
W^{\alpha\beta\rho\lambda}(p,q, C_T) = |C_T|^2\overline{\sum_{r,r'}} 
\langle H_c;  p',r' | \bar c(0) \sigma^{\alpha\beta} (1-\gamma_5) b(0) | H_b; p,r\rangle 
 \langle H_c; p',r' |\bar c(0) \sigma^{\rho\lambda} 
 (1-\gamma_5) b(0) | H_b; p,r \rangle^*, 
\ee
which contracted with  the lepton tensor $L_{\alpha\beta\rho\lambda}(k,k';h)$ 
in Eq.~\eqref{eq:leptensor} provides the contribution to the differential decay 
rate. Note that by construction 
$W^{\alpha\beta\rho\lambda}(p,q,C_T)= W^{\rho\lambda\alpha\beta* }(p,q,C_T)$, and hence, if
\bea
 W^{\alpha\beta\rho\lambda}&=&\frac12[ W^{\alpha\beta\rho\lambda}+
 W^{\rho\lambda\alpha\beta}]+\frac12[W^{\alpha\beta\rho\lambda}-
 W^{\rho\lambda\alpha\beta}] \nonumber \\
 &=& \frac12[ W^{\alpha\beta\rho\lambda}+
 W^{\alpha\beta\rho\lambda*}]+\frac12[W^{\alpha\beta\rho\lambda}-
 W^{\alpha\beta\rho\lambda*}]\equiv W^{\alpha\beta\rho\lambda}_{(s)}
 +W^{\alpha\beta\rho\lambda}_{(a)},
\eea 
the symmetric and antisymmetric, under the $(\alpha\beta) \leftrightarrow (\rho\lambda)$ exchange, 
parts become real and purely imaginary, respectively.
Now introducing the decomposition ($T=\sigma$, $pT=\sigma\gamma_5$)
\be
W^{\alpha\beta\rho\lambda}(p,q, C_T) = |C_T|^2 \left[
W^{\alpha\beta\rho\lambda}_{TT}(p,q) +  
W^{\alpha\beta\rho\lambda}_{pTpT}(p,q) - W^{\alpha\beta\rho\lambda}_{TpT} (p,q)
-  W^{\alpha\beta\rho\lambda}_{pTT}(p,q) \right],
\ee
and using parity and time-reversal, as in Eqs.~\eqref{eq:PT1} 
and \eqref{eq:PT2}, we conclude that  $W_{TT}^{\alpha\beta\rho\lambda}$ and 
$W_{pTpT}^{\alpha\beta\rho\lambda}$ ($W_{TpT}^{\alpha\beta\rho\lambda}$ and 
$W_{pTT}^{\alpha\beta\rho\lambda}$) are real  tensors 
(imaginary  pseudotensors).  Indeed,  we can identify
\bea
W^{\alpha\beta\rho\lambda}_{(s)} &=& |C_T|^2 \left[W^{\alpha\beta\rho\lambda}_{TT}(p,q) + 
W^{\alpha\beta\rho\lambda}_{pTpT}(p,q) \right],  \\
W^{\alpha\beta\rho\lambda}_{(a)} &=& -|C_T|^2 \left[W^{\alpha\beta\rho\lambda}_{TpT}(p,q) + 
W^{\alpha\beta\rho\lambda}_{pTT}(p,q) \right],
\eea
and conclude that the tensors/pseudotensors should be $(\alpha\beta) \leftrightarrow (\rho\lambda)$  symmetric/antisymmetric. In addition, both of them should be obviously antisymmetric under $\alpha\leftrightarrow \beta$ and   $\rho\leftrightarrow \lambda$ exchanges. There is still a large freedom, and a priori 5 (8) different four-index tensor (pseudotensor) structures, meeting all the above requirements, can be used  to construct  $W^{\alpha\beta\rho\lambda}_{(s)}$ ($W^{\alpha\beta\rho\lambda}_{(a)}$). An important simplification is found recalling that
\be
\gamma_5 \sigma^{\alpha\beta} = -\frac{i}{2} \epsilon^{\alpha\beta\delta\eta}
\sigma_{\delta\eta},\, \label{eq:sigma}
\ee
which can be used to relate  $W^{\alpha\beta\rho\lambda}_{(s)}$ and $W^{\alpha\beta\rho\lambda}_{(a)}$. We find
\be
W^{\alpha\beta\rho\lambda}= W^{\alpha\beta\rho\lambda}_{(s)}
-\frac{i}{2}\epsilon^{\rho\lambda}_{\ \ \delta\eta}W^{\alpha\beta\delta\eta}_{(s)},
\ee
which implies that the total tensor can be expressed using only the five real SFs 
that appear in the Lorentz decomposition\footnote{It is to say, they are defined from
\begin{eqnarray}
  \frac{W^{\alpha\beta\rho\lambda}_{(s)}}{|C_T|^2} &=& W^{\alpha\beta\rho\lambda}_{TT} +  W^{\alpha\beta\rho\lambda}_{pTpT}=  W_1^T 
  \left(g^{\alpha\rho}g^{\beta\lambda}-g^{\alpha\lambda}g^{\beta\rho}\right) + 
  \frac{W_2^T}{M^2} \left(g^{\alpha\rho}p^\beta p^\lambda-g^{\alpha\lambda}p^\beta p^\rho-
g^{\beta\rho}p^\alpha p^\lambda+g^{\beta\lambda}p^\alpha p^\rho\right)\nonumber \\ 
&&+ \frac{W_3^T}{M^2} \left(g^{\alpha\rho}q^\beta q^\lambda-g^{\alpha\lambda}q^\beta q^\rho-
g^{\beta\rho}q^\alpha q^\lambda+g^{\beta\lambda}q^\alpha q^\rho\right) + \frac{W_5^T}{M^4}
\left(p^\alpha q^\beta-p^\beta q^\alpha)(p^\rho q^\lambda-p^\lambda q^\rho\right)\nonumber\\
&&+ \frac{W_4^T}{M^2} \left(g^{\alpha\rho}(p^\beta q^\lambda+p^\lambda q^\beta)
-g^{\alpha\lambda}(p^\beta q^\rho+p^\rho q^\beta)-
g^{\beta\rho}(p^\alpha q^\lambda+p^\lambda q^\alpha)+g^{\beta\lambda}(p^\alpha q^\rho+p^\rho
q^\alpha)\right)
\end{eqnarray}
} of $W^{\alpha\beta\rho\lambda}_{(s)}$.
\bea
W^{\alpha\beta\rho\lambda} &= & |C_T|^2 \Bigg\{ W_1^T \Big[(g^{\alpha\rho}g^{\beta\lambda}-g^{\alpha\lambda}g^{\beta\rho})
-i\epsilon^{\rho\lambda\alpha\beta}\Big] + \frac{W_2^T}{M^2} \Big[(g^{\alpha\rho}p^\beta p^\lambda-g^{\alpha\lambda}p^\beta p^\rho-
g^{\beta\rho}p^\alpha p^\lambda+g^{\beta\lambda}p^\alpha p^\rho)\nonumber \\ 
&&-i\Big(
\epsilon^{\rho\lambda\alpha\delta}p^\beta p_\delta-
\epsilon^{\rho\lambda\beta\delta}p^\alpha p_\delta 
\Big)\Big] + \frac{W_3^T}{M^2} \Big[(g^{\alpha\rho}q^\beta q^\lambda-g^{\alpha\lambda}q^\beta q^\rho-
g^{\beta\rho}q^\alpha q^\lambda+g^{\beta\lambda}q^\alpha q^\rho) \nonumber\\
&&-i\Big(
\epsilon^{\rho\lambda\alpha\delta}q^\beta q_\delta-
\epsilon^{\rho\lambda\beta\delta}q^\alpha q_\delta\Big)\Big] + \frac{W_4^T}{M^2} \Big[\Big[g^{\alpha\rho}(p^\beta q^\lambda+p^\lambda q^\beta)
-g^{\alpha\lambda}(p^\beta q^\rho+p^\rho q^\beta)-
g^{\beta\rho}(p^\alpha q^\lambda+p^\lambda q^\alpha) \nonumber \\
&&+g^{\beta\lambda}(p^\alpha q^\rho+p^\rho q^\alpha)\Big]-i\Big(
\epsilon^{\rho\lambda\alpha\delta}(p^\beta q_\delta+q^\beta p_\delta)-
\epsilon^{\rho\lambda\beta\delta}(p^\alpha q_\delta+q^\alpha p_\delta)
\Big)\Big] \nonumber\\
&& + \frac{W_5^T}{M^4}\Big[(p^\alpha q^\beta-p^\beta q^\alpha)(p^\rho q^\lambda-p^\lambda q^\rho)
-i(p^\alpha q^\beta-p^\beta q^\alpha)\epsilon^{\rho\lambda\delta\eta}p_\delta
q_\eta \Big]\Bigg\}. \label{eq:WTdecom}
\eea
Requiring now that the pseudotensor part of $W^{\alpha\beta\rho\lambda}$  should be
 $(\alpha\beta) \leftrightarrow (\rho\lambda)$  antisymmetric, we find further constrains 
 for the $W_{1,2,3,4,5}^T$ since they should satisfy
\bea
&& M^2 W_{1}^T\Big[\epsilon^{\rho\lambda\alpha\beta}+\epsilon^{\alpha\beta\rho\lambda}\Big]
+W_{2}^T\Big[\Big(\epsilon^{\rho\lambda\alpha\delta}p^\beta p_\delta-
\epsilon^{\rho\lambda\beta\delta}p^\alpha p_\delta
\Big)+\Big(
\epsilon^{\alpha\beta\rho\delta}p^\lambda p_\delta-
\epsilon^{\alpha\beta\lambda\delta}p^\rho p_\delta
\Big)\Big]\nonumber\\
&&+W_{3}^T\Big[\Big(
\epsilon^{\rho\lambda\alpha\delta}q^\beta q_\delta-
\epsilon^{\rho\lambda\beta\delta}q^\alpha q_\delta\Big)+\Big(
\epsilon^{\alpha\beta\rho\delta}q^\lambda q_\delta-
\epsilon^{\alpha\beta\lambda\delta}q^\rho q_\delta\Big)\Big]\nonumber\\
&&+W_{4}^T\Big[\Big(
\epsilon^{\rho\lambda\alpha\delta}(p^\beta q_\delta+q^\beta p_\delta)-
\epsilon^{\rho\lambda\beta\delta}(p^\alpha q_\delta+q^\alpha p_\delta)
\Big)+\Big(
\epsilon^{\alpha\beta\rho\delta}(p^\lambda q_\delta+q^\lambda p_\delta)-
\epsilon^{\alpha\beta\lambda\delta}(p^\rho q_\delta+q^\rho p_\delta)
\Big)
\Big] \nonumber \\
&&+W_{5}^T \Big[ (p^\alpha q^\beta-p^\beta q^\alpha)\epsilon^{\rho\lambda\delta\eta}p_\delta
q_\eta + (p^\rho q^\lambda-p^\lambda q^\rho)\epsilon^{\alpha\beta\delta\eta}p_\delta q_\eta\Big]
=0.
\eea
The above equation can be rewritten as
\be
\epsilon^{\alpha\beta\rho\lambda} \Big [ 2 M^2 W_1^T + p^2 W_2^T + q^2 W_3^T+ 
2(p\cdot q) W_4^T
\Big] + q^2 p^\rho \epsilon^{\alpha\beta\delta\lambda}p_\delta \, W_5^T =0, \label{eq:WT-rest}
\ee
where we have used that
\be
\epsilon^{\alpha\beta\rho\delta}a_\delta b^\lambda- \epsilon^{\alpha\beta\lambda\delta}a_\delta
b^\rho + \epsilon^{\rho\lambda\alpha\delta}a_\delta b^\beta -
\epsilon^{\rho\lambda\beta\delta}a_\delta b^\alpha = (a\cdot b)
\epsilon^{\alpha\beta\rho\lambda}.
\ee
Taking into account that the two tensors that appear  in Eq.~\eqref{eq:WT-rest} are 
independent, we deduce 
\be
2 M^2 W_1^T + p^2 W_2^T + q^2 W_3^T+ 2(p\cdot q) W_4^T =0, \qquad W_5^T = 0. \label{eq:relacWTs}
\ee
The first of the above equations can be used to re-write $W_1^T$ in terms 
of $W^T_{2,3,4}$. Nevertheless, the contraction of the tensor that multiplies  $W_1^T$ in 
the decomposition of Eq.~\eqref{eq:WTdecom} with the lepton tensor  $L_{\alpha\beta\rho\lambda}(k,k';h)$ defined 
 in Eq.~\eqref{eq:leptensor} vanishes identically. Hence, the contribution of 
 $ W^{\alpha\beta\rho\lambda}$ to $\overline\sum  |{\cal M}|^2$ 
 is given just in terms of three ($W_2^T$, $W_3^T$, $W_4^T$) real SFs. 

 Finally, we  absorb the common factor $|C_T|^2$, by redefining 
 $\widetilde W^T_{1,2,3,4} = |C_T|^2 
W^T_{1,2,3,4}$. 

\item  The diagonal contribution of the  operator $O_H$ gives rise to the scalar
\be
W(p,q)= \widetilde W_{SP}(q^2) = |C_S|^2 \overline{\sum_{r,r'}} |\langle H_c;  p',r' | \bar c(0)
 b(0) | H_b; p,r\rangle|^2 + |C_P|^2 \overline{\sum_{r,r'}} |\langle H_c;  p',r' | \bar c(0)
 \gamma_5 b(0) | H_b; p,r\rangle|^2,  
\ee
which should be multiplied by the scalar lepton term of Eq.~\eqref{eq:leptensorSP}. Note that the $C_S C_P^*$ and $C_P C_S^*$ interference terms would give rise to purely imaginary pseudoscalars, which necessarily  vanish because they cannot be constructed out of  the $p$ and $q$ four-vectors alone.  
\item The $O_H^\alpha$ and $O_H$ interference contribute to the decay width as 
$2{\rm Re}\left[W^{\alpha}(p,q, C_{V,A,S,P})L_\alpha(k,k';h)\right]$, with 
the $L_\alpha(k,k';h)$ lepton tensor defined in Eq.~\eqref{eq:leptensor-vec} and 
 \be W^{\alpha}(p,q, C_{V,A,S,P}) = \overline{\sum_{r,r'}} \langle H_c;  p',r' | (C_V V^\alpha- C_A A^\alpha) | H_b; p,r\rangle 
 \langle H_c; p',r' |\bar c(0) (C_S- C_P\gamma_5) b(0) | H_b; p,r \rangle^*,
\ee
and its treatment is similar to that discussed for $J_{H}^\alpha [J_{H}^\rho]^*$ in 
Sec.~\ref{sec:hme}, with the equivalence $(VV) \leftrightarrow (VS)$, $(AA) \leftrightarrow (AP)$, $(V\!A) \leftrightarrow (VP)$ and $(AV) \leftrightarrow (AS)$. Thus, we find
\bea 
W^{\alpha}(p,q, C_{V,A,S,P}) &=& \frac{1}{2M} \left( \widetilde  W_{I1} p^\alpha +\widetilde 
W_{I2} q^\alpha \right), \\
\widetilde W_{I1, I2} (q^2, C_{V,A,S,P})&=& C_V C^*_S W^{VS}_{I1,I2}(q^2) + C_A C^*_P
W^{AP}_{I1,I2}(q^2),
\eea
with all four $W^{VS,AP}_{I1,I2}$ SFs being real, and where we have used an obvious notation in 
which $W_{I1,I2}^{VS}$,  and $W_{I1,I2}^{AP}$
should be obtained from the $VS$ and $AP$ matrix elements. Note that the odd parity $VP$ 
and $AS$ terms would give rise to purely imaginary pseudo-vectors, which necessarily  vanish 
because they cannot be constructed out of  $p$ and $q$ alone. Thus, the total contribution 
to $\overline\sum  |{\cal M}|^2$ of these pieces is given by
\be
  {\rm Re}\left[\left(\frac{\widetilde W_{I1}}{M} p^\alpha + \frac{\widetilde  W_{I2}}{M}
  q^\alpha\right) L_\alpha(k,k';h)\right]. \label{eq:contr-vec}
\ee
For real Wilson coefficients, the $\widetilde W_{I1,I2}$ SFs are real, and taking  the real part 
in Eq.~\eqref{eq:contr-vec} amounts to remove the Levi-Civita term of $L_\alpha(k,k';h)$, 
recovering in this way the result of Ref.~\cite{Penalva:2019rgt} identifying 
$\widetilde W_{I1,I2}$ with 
$W_{I1,I2}$ introduced in the latter reference.

\item The $O_H$ and $O_H^{\rho\lambda}$ interference contribute to the decay 
width as $2{\rm Re}\left[W^{\prime\rho\lambda}(p,q, C_{S,P,T})L'_{\rho\lambda}(k,k';h)\right]$,
 with the $L'_{\rho\lambda}(k,k';h)$ lepton tensor defined in Eq.~\eqref{eq:leptensor-ST} and 
\be 
 W^{\prime\rho\lambda}(p,q, C_{S,P,T}) = C_T^*\overline{\sum_{r,r'}} \langle H_c;  p',r' |
  \bar c(0) (C_S - C_P \gamma_5) b(0) | H_b; p,r\rangle 
 \langle H_c; p',r' |\bar c(0) \sigma^{\rho\lambda}(1-\gamma_5) b(0) | H_b; p,r \rangle^*.
\ee
We use Lorentz, parity and time-reversal transformations, as explained in Eqs.~\eqref{eq:PT1} and \eqref{eq:PT2}, to deduce that the $ST$ and $PpT$ 
($SpT$ and  $PT$)  tensors are purely imaginary (real) antisymmetric  
tensors (pseudotensors). In addition,  the $SpT$ and $PT$ pseudotensors can 
be  related to the $ST$ and $PpT$ tensors thanks to Eq.~\eqref{eq:sigma}. 
We finally find 
\bea
 W^{\prime\rho\lambda}(p,q, C_{S,P,T}) &=&  \frac{\widetilde W_{I3}}{2M^2} \left[ 
 \epsilon^{\rho\lambda\delta\eta} p_\delta q_\eta + i(p^\rho q^\lambda-p^\lambda q^\rho) 
 \right], \nonumber \\
 \widetilde W_{I3}(q^2, C_{S,P,T}) &=& C_T^* \left(C_S W_{I3}^{ST}(q^2)+ C_P
 W_{I3}^{PpT}(q^2)\right),
\eea
and the real $W_{I3}^{ST}$ and $W_{I3}^{PpT}$ SFs, obviously,  deduced from the decompositions
\bea
 W^{\prime PpT}_{\rho\lambda}(p,q) &=& \overline{\sum_{r,r'}} \langle H_c;  p',r' | 
 \bar c(0) \gamma_5 b(0)| H_b; p,r\rangle 
 \langle H_c; p',r' |\bar c(0) \sigma_{\rho\lambda}\gamma_5 b(0) | H_b; p,r \rangle^* = 
 \frac{W_{I3}^{PpT}}{2M^2}\,i(p_\rho q_\lambda-p_\lambda q_\rho), \nonumber \\
 W^{\prime ST}_{\rho\lambda}(p,q) &=& \overline{\sum_{r,r'}} \langle H_c;  p',r' | 
 \bar c(0) b(0)| H_b; p,r\rangle 
 \langle H_c; p',r' |\bar c(0) \sigma_{\rho\lambda} b(0) | H_b; p,r \rangle^* =
 \frac{W_{I3}^{ST}}{2M^2}\,i(p_\rho q_\lambda-p_\lambda q_\rho). 
\eea
Its total contribution to $\overline\sum  |{\cal M}|^2$ is given by
\be
  {\rm Re}\left\{\frac{\widetilde W_{I3}}{M^2} \left[\epsilon^{\rho\lambda\delta\eta} 
  p_\delta q_\eta + i(p^\rho q^\lambda-p^\lambda q^\rho) \right] L'_{\rho\lambda}(k,k';h)\right\} 
\ee

\item The $O^\alpha_H$ and $O_H^{\rho\lambda}$ interference contribute to the decay width 
as $2{\rm Re}\left[W^{\alpha\rho\lambda}(p,q, C_{V,A,T})L_{\alpha\rho\lambda}(k,k';h)
\right]$, with the $L_{\alpha\rho\lambda}(k,k';h)$ lepton tensor defined in Eq.~\eqref{eq:lepdif} and 
\be 
 W^{\alpha\rho\lambda}(p,q, C_{V,A,T}) = C_T^*\overline{\sum_{r,r'}} \langle H_c;  p',r' | 
 C_V V^\alpha-C_A A^\alpha | H_b; p,r\rangle 
 \langle H_c; p',r' |\bar c(0) \sigma^{\rho\lambda}(1-\gamma_5) b(0) | H_b; p,r \rangle^*.
\ee
The analysis runs in parallel to the  previous one for 
$J_{H} [J_{H}^{\rho\lambda}]^*$, identifying $V$ and $A$ here with $S$ 
and $P$ that appeared 
previously. The only difficulty is that now there are four, instead of one, 
independent Lorentz structures. We use parity and time-reversal 
transformations to deduce that the $VT$ and $ApT$ ($VpT$ and  $AT$)  tensors are purely imaginary (real)  tensors (pseudotensors), and obviously antisymmetric under the $\rho \leftrightarrow \lambda$ exchange. Here again,  the $VpT$ and $AT$ pseudotensors can be  related to the $VT$ and $ApT$ tensors thanks to Eq.~\eqref{eq:sigma}, and we finally find
\bea
 W^{\alpha\rho\lambda}(p,q, C_{V,A,T}) &=&  \frac{p^\alpha \widetilde  W_{I4} + q^\alpha \widetilde W_{I5}}{2M^3} 
 \left[ \epsilon^{\rho\lambda\delta\eta} p_\delta q_\eta + i(p^\rho q^\lambda-p^\lambda q^\rho) 
 \right] \nonumber \\
&&+ \frac{p_\delta \widetilde  W_{I6} + q_\delta \widetilde W_{I7}}{2M} \left[ \epsilon^{\rho\lambda\alpha\delta}  +
 i(g^{\alpha\rho}g^{\lambda\delta}-g^{\alpha\lambda}g^{\rho\delta})\right],\\
 \widetilde W_{I4,I5,I6,I7}(q^2, C_{V,A,T}) &=& C_T^* \left(C_V W_{I4,I5,I6,I7}^{VT}(q^2)+ C_A
 W_{I4,I5,I6,I7}^{ApT}(q^2)\right),
\eea
and the real $W_{I4,I5,I6,I7}^{VT}$ SFs are  deduced from 
\bea
 W^{ VT}_{\alpha\rho\lambda}(p,q) &=& \overline{\sum_{r,r'}} \langle H_c;  p',r' | 
 V_\alpha| H_b; p,r\rangle 
 \langle H_c; p',r' |\bar c(0) \sigma_{\rho\lambda} b(0) | H_b; p,r \rangle^* \nonumber \\
 &=& \frac{p_\alpha  W_{I4}^{VT} + q_\alpha  W_{I5}^{VT}}{2M^3} i(p_\rho q_\lambda
 -p_\lambda q_\rho)  
+ \frac{p^\delta  W_{I6}^{VT} + q^\delta  W_{I7}^{VT}}{2M} 
 i(g_{\alpha\rho}g_{\lambda\delta}-g_{\alpha\lambda}g_{\rho\delta}), 
 \eea
while $W_{I4,I5,I6,I7}^{ApT}$ are obtained from a similar decomposition replacing $V^\alpha$
 by $A^\alpha$ and $\left[\bar c(0) \sigma_{\rho\lambda} b(0)\right]$ by 
$\left[\bar c(0) \sigma_{\rho\lambda}\gamma_5 b(0)\right]$. The total contribution to 
$\overline\sum  |{\cal M}|^2$ of is given by
\bea
  &&{\rm Re}\Bigg\{ \Big[\frac{p^\alpha \widetilde  W_{I4} + q^\alpha \widetilde W_{I5}}{M^3}
   \left[ \epsilon^{\rho\lambda\delta\eta} p_\delta q_\eta + i(p^\rho q^\lambda-p^\lambda 
   q^\rho) \right] \nonumber \\
&&\hspace{1cm}+ \frac{p_\delta \widetilde  W_{I6} + q_\delta \widetilde W_{I7}}{M} \left[ \epsilon^{\rho\lambda\alpha\delta}  +
 i(g^{\alpha\rho}g^{\lambda\delta}-g^{\alpha\lambda}g^{\rho\delta})\right]\Big]
  L_{\alpha\rho\lambda}(k,k';h)\Bigg\} .
\eea
\end{itemize}
\section{CM and LAB kinematics}
\label{sec:scalaP}

To compute the contractions of the lepton and hadron tensors, we use
\begin{equation}
 p^2=M^2,\,\, k^2=0,\,\, k^{\prime 2}= m^2_\ell,\,\, p\cdot q= MM_\omega, 
 \,\, s^2= -m^2_\ell, \,\,k'\cdot s=0,\,\, k\cdot k'= q\cdot k=\frac{q^2-m^2_\ell}{2}, 
 \,\, q\cdot k'=\frac{q^2+m^2_\ell}{2},
\end{equation}
with $M_\omega=M-M'\omega$. In addition, the scalar products that depend
 explicitly on the charged lepton variables used in the differential decay
 widths  read 
\begin{itemize}
 \item CM
\begin{eqnarray}
 p\cdot k= \frac{M}{2} \left(1-\frac{m^2_\ell}{q^2}\right)\left(M_\omega + M'\sqrt{\omega^2-1}
 \cos\theta_\ell\right),\,\,  k\cdot s= \frac{q^2-m^2_\ell}{2},\,\, p\cdot s = 
 MM_\omega-p\cdot k \, \frac{q^2+m_\ell^2}{q^2-m_\ell^2}.
\end{eqnarray}
 \item LAB
\begin{eqnarray}
 p\cdot k= M(M_\omega-E_\ell), \,\, k\cdot s= \frac{E_\ell(q^2+m^2_\ell)-2m^2_\ell M_\omega}
 {2(E_\ell^2-m^2_\ell)^\frac12},\,\, p\cdot s = 
 M(E_\ell^2-m_\ell^2)^\frac12.
\end{eqnarray}
\end{itemize}
Note that both in the CM and LAB frames, $\epsilon_{\delta\eta\mu\nu}k^\delta 
q^\eta s^\mu p^\nu=0$, which trivially follows from 
$s^\mu = \frac{k^{\prime 0}}{|\vec{k}'\,|} k^{\prime \mu} 
-\frac{m^2_\ell}{|\vec{k}'\,|} n^\mu $ with $n^\mu=(1,\vec{0}\,)$, $q=k+k'$ and the fact that $p^\mu_{\rm LAB}$  and $q^\mu_{\rm CM}$ are proportional to $n^\mu$. 

\section{ Coefficients of the CM  $A(\omega,\theta_\ell)$  and  LAB $C(\omega,E_\ell)$ distributions in terms of the $\widetilde W$ SFs}
\label{app:coeff}

In this appendix we collect the expressions of the 
${\cal A}, {\cal B}$ and ${\cal C}$ functions as well as the ${\cal A}_H,{\cal B}_H,{\cal C}_H,
{\cal D}_H$ and ${\cal E}_H$ functions that together determine the expansion 
coefficients of the  CM  $A_h(\omega,\theta_\ell)$  and  
LAB $C_h(\omega,E_\ell)$ distributions   
(Eqs.~\eqref{eq:aspol}-\eqref{eq:cespol}). They are combinations of the hadronic $\widetilde W$  
SFs and are given by
\begin{eqnarray}
 {\cal A}(\omega) &=&\frac{q^2-m_\ell^2}{M^2}\left\{ 2\widetilde W_1 -
 \widetilde W_2+\frac{M_\omega}{M}\widetilde W_3+ \widetilde W_{SP} 
 + \frac{4M_\omega}{M} {\rm Re}[\widetilde W_{I3}]
 +8 \left(\widetilde W_2^T-\frac{q^2}{M^2}\widetilde W_3^T- \frac{2 M_\omega}{M}
 \widetilde W_4^T\right)\right.\nonumber \\
 &&\hspace{2cm}\left.+ \frac{m_\ell}{M}{\rm Re}\left[\widetilde W_{I2}+4\,\widetilde W_{I4}+
 \frac{4M_\omega}{M} \widetilde W_{I5}+12\,\widetilde W_{I7}\right]+
 \frac{m_\ell^2}{M^2}\left( \widetilde W_4-16 \widetilde W_3^T\right)\right\},\nonumber \\
 {\cal B}(\omega) & = & -\frac{2q^2}{M^2}\left(\widetilde W_3+ 
 4{\rm Re}[ \widetilde W_{I3}]\right)+\frac{4M_\omega}{M}
 \Big(\widetilde W_2 -16\, \widetilde W_2^T \Big)\nonumber
\\&& +  \frac{2m_\ell}{M}{\rm Re}\left[\widetilde W_{I1} 
 -\frac{4 M_\omega}{M}\widetilde W_{I4}-\frac{ 4 q^2}{M^2}\widetilde W_{I5}+12\,
 \widetilde W_{I6}\right] +\frac{2m_\ell^2}{M^2} \left( \widetilde W_5 -32\, 
 \widetilde W_4^T\right), \nonumber \\
 {\cal C}(\omega)&=& -4\Big(\widetilde W_2 -16\, \widetilde W_2^T \Big),\label{eq:abc}
\end{eqnarray}
\begin{eqnarray}
{\cal A_H}(\omega)&=& -\frac{q^2-m^2_\ell}{2M^2}\Bigg\{{\rm Re} \left[\widetilde W_{I1}+\frac{4 M_{\omega}}{M} \widetilde W_{I4}-4\,\widetilde W_{I6} \right]
+\frac{m_\ell}{M}\left( \widetilde W_3 +\widetilde W_5 -4\,{\rm Re}[\widetilde W_{I3}]+ 32\, \widetilde W_4^T   \right)
-\frac{4 m_\ell^2}{M^2} {\rm Re} [\widetilde W_{I5}]\Bigg\},
\nonumber\\
\nonumber\\
{\cal B_H}(\omega)&=& \frac{M_\omega}{M}{\rm Re} \left[\widetilde W_{I1}+\frac{4 M_{\omega}}{M} \widetilde W_{I4}-4\,\widetilde W_{I6} \right] 
-\frac{m_\ell}{M}\left(2\, \widetilde W_1 - \widetilde W_2  - \frac{M_\omega}{M}\widetilde W_5 - \widetilde W_{SP} - 8\, \widetilde W_2^T +\frac{8q^2}{M^2}\widetilde W_3^T - \frac{16 M_\omega}{M} \widetilde W_4^T  \right)\nonumber \\
&&+ \frac{m_\ell^2}{M^2} {\rm Re} \left[\widetilde W_{I2}-4\,\widetilde W_{I4}-4\,
\widetilde W_{I7}\right]+ \frac{m_\ell^3}{M^3}\left(\widetilde W_4 + 16\, 
\widetilde W_3^T \right), \nonumber \\ \nonumber \\
{\cal C_H}(\omega)&=& \frac{4q^2}{M^2}{\rm Re}[\widetilde W_{I4}]-\frac{2m_\ell}{M}
\left(\widetilde W_2+16 \widetilde W_2^T\right)  + \frac{4 m_\ell^2}{M^2}{\rm Re}[\widetilde
W_{I4}], \nonumber \\\nonumber \\
{\cal D_H}(\omega)&=& - {\rm Re} \left[\widetilde W_{I1}+\frac{12 M_\omega}{M} 
\widetilde W_{I4}-4\, \widetilde W_{I6}\right] + \frac{m_\ell}{M}\left(
\widetilde W_3 - \widetilde W_5- 32 \widetilde W_4^T  - 4 {\rm Re}
[\widetilde W_{I3}]\right) - \frac{4m^2_\ell}{M^2} {\rm Re}[\widetilde W_{I5}], \nonumber \\\nonumber \\
{\cal E_H}(\omega)&=& 8\, {\rm Re}[\widetilde W_{I4}].\label{eq:polgral2}
\end{eqnarray}
Using the above relations and Eqs.~(\ref{eq:aspol}) and  (\ref{eq:cespol}) 
one can obtain explicit expressions in terms of the $\widetilde W$  
SFs for the $a_{1,2,3}(\omega,h)$ and $\widehat c_{0,1,2,3}$ expansion coefficients.
They are given by
\begin{eqnarray}
a_0(h=+1) &=& \frac{8q^2}{M^2}\left(\frac{\widetilde W_ {SP}}{8}+ 
\widetilde W_2^T-\frac{q^2}{M^2}\widetilde W_3^T-\frac{2 M_\omega}{M}\widetilde W_4^T\right) 
-\frac{16 M_\omega^2}{M^2}W_2^T \nonumber \\
 &&+\frac{m_\ell}{M}{\rm Re}\left[\frac{M_\omega}{M}\widetilde W_{I1}
 +\frac{q^2}{M^2}\widetilde W_{I2}+\frac{4M_\omega}{M}\widetilde W_{I6}
 +\frac{4q^2}{M^2}\widetilde W_{I7}\right]+\frac{m_\ell^2}{M^2}\left( \frac{M_\omega^2}{q^2}
 \widetilde W_2+\frac{q^2}{M^2} \widetilde W_4+\frac{M_\omega}{M}\widetilde W_5\right), \nonumber \\
 a_1(h=+1) &=&\sqrt{\omega^2-1}\,\frac{M'}{M}\Bigg\{- 4\frac{q^2}{M^2}{\rm Re}[\widetilde W_{I3}]+\frac{4m_\ell}{M}{\rm Re}\left[\frac{\widetilde W_{I1}}{4}-\frac{M_\omega}{M}\widetilde W_{I4} -\frac{q^2}{M^2}\widetilde W_{I5}+\widetilde W_{I6}\right]\nonumber\\
 &&\hspace{2.5cm}+\frac{m_\ell^2}{M^2} \left ( \frac{2M M_\omega}{q^2}\widetilde W_2
 +\widetilde W_5\right) \Bigg\},\nonumber \\
 a_2(h=+1)&=& (\omega^2-1)\frac{M'^2}{M^2} \left( 16\widetilde W_2^T  
 -4\frac{m_\ell}{M} {\rm Re}[\widetilde W_{I4}]+\frac{m^2_\ell}{q^2}\widetilde W_2\right), 
\label{eq:aesplus1}
 \end{eqnarray}
\begin{eqnarray}
 a_0(h=-1) &=& \frac{2q^2}{M^2}\widetilde W_1 -\frac{q^2-M^2_\omega}{M^2}\widetilde W_2+ \frac{4m_\ell}{M}{\rm Re}\left[ \frac{q^2-M_\omega^2}{M^2}\widetilde W_{I4}+\frac{2M_\omega}{M}\widetilde W_{I6}+\frac{2q^2}{M^2}\widetilde W_{I7}\right]
\nonumber\\
&&- \frac{16m^2_\ell}{M^2}\left(\frac{M_\omega^2}{q^2}\widetilde W_2^T 
+  \frac{q^2}{M^2}\widetilde W_3^T +\frac{2 M M_\omega}{M^2}\widetilde W_4^T  \right),\nonumber\\
a_1(h=-1) &=& -\sqrt{\omega^2-1}\,\frac{M'}{M}\Bigg\{\frac{q^2}{M^2}\widetilde W_3
- \frac{8m_\ell}{M}{\rm Re}[\widetilde W_{I6}]+\frac{32m_\ell^2}{M^2}\left( 
\frac{M M_\omega}{q^2}\widetilde W_2^T+ \widetilde W_4^T\right) \Bigg\},\nonumber \\
a_2(h=-1) &=&-(\omega^2-1)\frac{M'^2}{M^2} 
\left(\widetilde W_2-\frac{4m_\ell}{M}{\rm Re}[\widetilde W_{I4}]
+\frac{16m_\ell^2}{q^2} \widetilde W_2^T\right),\label{eq:aesmenos1}\\ \nonumber\\\nonumber\\
 \widehat{c}_0(\omega) & = & \frac{4q^2}{M^2}\frac{m_\ell}{M}{\rm Re}\left[
 - \frac{\widetilde W_{I1}}{4}+ \frac{M_\omega}{M} \widetilde W_{I4}+  
 \widetilde W_{I6}\right] 
 \nonumber\\&&-\frac{m_\ell^2}{M^2}\Bigg \{\frac{q^2}{M^2} \left(\widetilde W_3 
 +\widetilde W_5-4 {\rm Re}[\widetilde W_{I3}]+\frac{16M_\omega}{M} \widetilde W_3^T+32\, 
 \widetilde W_4^T\right ) +  \frac{2M_\omega}{M}\left( 2 \widetilde W_1+\widetilde W_2-\widetilde W_{SP}+24 \widetilde W_2^T \right)
\nonumber \\
 &&\hspace{1.5cm} -\frac{8M_\omega^2}{M^2} \left(\frac{\widetilde W_3}{4}- {\rm Re}[\widetilde W_{I3} ]-4 \widetilde W_4^T\right)\Bigg\} \nonumber \\ 
 && +\frac{2m_\ell^3}{M^3}{\rm Re}\left[\frac{\widetilde W_{I1}}{2}+\frac{M_\omega}{M} 
 \left(\widetilde W_{I2}+2 \widetilde W_{I4}-4 
 \widetilde W_{I7}\right)+ \frac{2q^2-4M_\omega^2}{M^2}\, \widetilde W_{I5}
-2\, \widetilde W_{I6}\right] \nonumber \\
&&+\frac{m_\ell^4}{M^4}\left (\widetilde W_3+  \frac{2 M_\omega}{M} 
\left(\widetilde W_4+ 16 \widetilde W_3^T\right )+\widetilde W_5  
-4 {\rm Re}[\widetilde W_{I3}]+ 32\, \widetilde W_4^T\right) 
-\frac{4  m_\ell^5}{M^5}{\rm Re}[\widetilde W_{I5}],
\nonumber\\
 \widehat{c}_1(\omega) & = & \frac{8 q^2}{M^2} \left(-\frac{\widetilde W_{SP}}{4}
 +\frac{M_\omega}{M} \left({\rm Re}[\widetilde W_{I3}]+4 \widetilde W_4^T \right)
 - 2 \widetilde W_2^T + \frac{2q^2}{M^2}\widetilde W_3^T \right)
 \nonumber \\
 &&-\frac{8m_\ell}{M} {\rm Re} \left[ \frac{q^2}{M^2}\left(
\frac{\widetilde W_{I2}}{4}+ \widetilde W_{I4}- \frac{M_\omega}{M} 
\widetilde W_{I5}+\widetilde W_{I7}\right) + \frac{4M_\omega}{M} \widetilde W_{I6}\right]
\nonumber\\&&+  \frac{m_\ell^2}{M^2}\Bigg\{ 4\, \widetilde W_1+2\,\widetilde W_2+64\,
\widetilde W_2^T  -\frac{16M_\omega}{M} \left(\frac{\widetilde W_3}{8}-8 \widetilde W_4^T 
- {\rm Re}[\widetilde W_{I3}]\right)- \frac{2q^2}{M^2}\widetilde W_4 \Bigg\}
\nonumber \\
&&+\frac{16 m_\ell^3}{M^3}{\rm Re}\left[\frac{M_\omega}{M} \widetilde W_{I5}
+ \widetilde W_{I7}\right] -\frac{32m_\ell^4}{M^4} \widetilde W_3^T,  
\nonumber\\
  \widehat{c}_2(\omega) & = & -128 \left(\frac{q^2}{8M^2}{\rm Re}
  [\widetilde W_{I3}]-\frac{M_\omega}{M} \widetilde W_2^T \right)
   -\frac{16 m_\ell}{M} {\rm Re}\left[ \frac{q^2}{M^2}
\widetilde W_{I5}-2 \widetilde W_{I6}\right] - \frac{128 m_\ell^2}{M^2}
\widetilde W_4^T, \nonumber \\
  \widehat{c}_3(\omega) & = &-128\,\widetilde W_2^T. \label{eq:cesconpol}
\end{eqnarray}

  \section{Hadron tensor SFs and form-factors for the $\Lambda_b^0\to  \Lambda_c^+\ell^-\bar\nu_\ell$ decay}
\label{app:ff}
The form factors used in Eq.~\eqref{eq.FactoresForma} are  related to those computed in Refs.~\cite{Detmold:2015aaa} and \cite{Datta:2017aue} by
\begin{eqnarray}
F_1&=&f_{\perp},\quad G_1 =g_{\perp},\quad F_S = \frac{\delta_{M_\Lambda}}{m_b-m_c}f_0,\quad
F_P= \frac{\Delta_{M_\Lambda}}{m_b+m_c}g_0,\quad T_4=\widetilde h_+,\nonumber\\
F_2&=&\frac{M_{\Lambda_b}\delta_{M_\Lambda}}{q^2}f_0+
\frac{M_{\Lambda_b}\Delta_{M_\Lambda}}{s_+}
\left[1-\delta\right]f_+-\delta_{s_{+}}f_{\perp},\nonumber\\
F_3&=&-\frac{M_{\Lambda_c}\delta_{M_\Lambda}}{q^2}f_0
+\frac{M_{\Lambda_c}\Delta_{M_\Lambda}}{s_+}
\left[1+\delta\right]f_+-\delta_{s_{+}}f_{\perp},\nonumber \\
G_2&=&-\frac{M_{\Lambda_b}\Delta_{M_\Lambda}}{q^2}g_0-\frac{M_{\Lambda_b}\delta_{M_\Lambda}}{s_-}\left[1-\delta\right]g_+
-\delta_{s_{-}}g_{\perp},\nonumber\\
G_3&=&\frac{M_{\Lambda_c}\Delta_{M_\Lambda}}{q^2}g_0-\frac{M_{\Lambda_c}\delta_{M_\Lambda}}{s_-}\left[1+\delta\right]g_+
+\delta_{s_{-}}g_{\perp},\nonumber\\
\nonumber\\
T_1 &=& -2M^2\Big[ \frac{h_+}{s_+}-
\frac{\Delta_{M_\Lambda}^2}{q^2 s_+}h_\perp -\frac{\widetilde
h_+}{s_-}+\frac{\delta_{M_\Lambda}^2}{q^2\,s_-}\,\widetilde h_\perp\Big],\nonumber\\
T_2&=&M\Big[\frac{\Delta_{M_\Lambda}}{q^2}\, h_\perp+\frac{2M_{\Lambda_c}}{s_-}\,\widetilde h_+
+\frac{\delta_{M_\Lambda}(1-\delta)}{s_-}\,\widetilde h_\perp\Big],\nonumber\\
T_3&=&M\Big[-\frac{\Delta_{M_\Lambda}}{q^2}\, h_\perp-\frac{2M_{\Lambda_b}}{s_-}\,\widetilde
h_+ +\frac{\delta_{M_\Lambda}(1+\delta)}{s_-}\,\widetilde h_\perp\Big],\nonumber\\
\end{eqnarray}
with $\delta=(M_{\Lambda_b}^2-M_{\Lambda_c}^2)/q^2$, $s_{\pm}=(M_{\Lambda_b}\pm M_{\Lambda_c})^2-q^2$, 
$\delta_{M_\Lambda}=M_{\Lambda_b}-M_{\Lambda_c}$, $\Delta_{M_\Lambda}=M_{\Lambda_b}+M_{\Lambda_c}$ and $\delta_{s_{\pm}}=
2M_{\Lambda_b}M_{\Lambda_c}/s_{\pm}$. Note that $F_S$ and $F_P$ have not been computed in LQCD, and both form-factors are obtained from the vector $f_0$ and axial $g_0$ form factors using the equations of motion. In the numerical calculations, we use  $m_b = 4.18\pm 0.04$ GeV and  
$m_c = 1.27 \pm 0.03$ GeV as in Ref.~\cite{Datta:2017aue}. For completeness, these two latter form factors  are related to those introduced 
in Eq.~\eqref{eq.FactoresForma}  by
\begin{eqnarray}
 f_0 &=& F_1 + \frac{F_2 \left(M_{\Lambda_b}-\omega M_{\Lambda_c}\right)+ F_3\left(\omega M_{\Lambda_b}- M_{\Lambda_c}\right)}{M_{\Lambda_b}- M_{\Lambda_c}} \nonumber \\
 g_0 &=& G_1 -\frac{G_2 \left(M_{\Lambda_b}-\omega M_{\Lambda_c}\right)+ G_3\left(\omega M_{\Lambda_b}- M_{\Lambda_c}\right)}{M_{\Lambda_b}+ M_{\Lambda_c}}
\end{eqnarray}
and in the heavy quark limit $f_0=g_0=\zeta + {\cal O}\left(\alpha_s, \Lambda_{\rm QCD}/m_{c,b} \right)$. 

On the other hand, from Eqs.~\eqref{eq:wmunu-baryons} and the results of 
Appendix~\ref{sec:appH}, we find for the $\widetilde W$ SFs related to the SM currents 
 \begin{eqnarray}
\widetilde W_1&=& \frac{1}{2}\Big[(\omega-1)|C_V|^2 F_1^2+(\omega+1)|C_A|^2 G_1^2\Big],\nonumber\\
\widetilde W_2&=&\frac{|C_V|^2}{2}\left\{ 2F_1F_2+(\omega+1) F_2^2+\frac{2M_{\Lambda_b}}{M_{\Lambda_c}}\Big[(F_1+F_2)(F_1+F_3)+\omega F_2F_3\Big] 
+\frac{M_{\Lambda_b}^2}{M_{\Lambda_c}^2}\Big[2F_1F_3+(\omega+1)F_3^2\Big] \right\} + \nonumber \\
&& \frac{|C_A|^2}{2}\left\{2G_1G_2+(\omega-1)G_2^2
+\frac{2M_{\Lambda_b}}{M_{\Lambda_c}}
\Big[\omega G_2G_3 + (G_1-G_2)(G_1+G_3)\Big]+\frac{M_{\Lambda_b}^2}{M_{\Lambda_c}^2}\Big[
(\omega-1)G_3^2-2G_1G_3\Big]\right\},\nonumber\\
\widetilde W_3&=& \frac{2M_{\Lambda_b}}{M_{\Lambda_c}}{\rm Re}[C_V C_A^*]F_1G_1,\nonumber\\
\widetilde W_4&=& \frac{M_{\Lambda_b}^2}{2M_{\Lambda_c}^2}
\Big[ |C_V|^2 \left( 2F_1F_3+(\omega+1)F_3^2\right)-|C_A|^2 \left( 2G_1G_3+(1-\omega)G_3^2\right)\Big],\nonumber\\
\widetilde W_5&=&-\frac{M_{\Lambda_b}}{M_{\Lambda_c}}|C_V|^2 \left[(F_1+F_2)(F_1+F_3)+\omega F_2F_3 +\frac{M_{\Lambda_b}}{M_{\Lambda_c}}\Big[2F_1F_3 +(\omega+1)F_3^2 \Big]
\right]\nonumber\\
&&-\frac{M_{\Lambda_b}}{M_{\Lambda_c}}|C_A|^2 \left[(G_1-G_2)(G_1+G_3) +\omega G_2G_3 
-\frac{M_{\Lambda_b}}{M_{\Lambda_c}}\Big[2G_1G_3+ (1-\omega)G_3^2\Big]\right]
\end{eqnarray}
The rest of NP $\widetilde W$ SFs for this baryon decay are
\begin{eqnarray}
\widetilde W_{SP}&=& \frac{1}{2}\Big[(\omega+1) |C_S|^2  F_S^2+(\omega-1) |C_P|^2 F_P^2\Big],
\nonumber\\
\widetilde W_{I1}&=&  C_V C_S^* \left[F_S F_1 \left(1+\frac{M_{\Lambda_b}}{M_{\Lambda_c}}\right) + (1+\omega)F_S \left(F_2+\frac{M_{\Lambda_b}}{M_{\Lambda_c}}F_3\right)  \right]\nonumber \\
&& + C_A C_P^* \left[ F_P G_1\left(1-\frac{M_{\Lambda_b}}{M_{\Lambda_c}}\right) - (1-\omega) F_P \left( G_2 + \frac{M_{\Lambda_b}}{M_{\Lambda_c}} G_3\right) \right],\nonumber\\
\widetilde W_{I2}&=& \frac{M_{\Lambda_b}}{M_{\Lambda_c}}\Bigg\{ C_A C_P^* \,F_P\Big[G_1+ (1-\omega)G_3 \Big]- C_V C_S^* \, F_S\Big[F_1+ (1+\omega)F_3 \Big]\Bigg\}, 
\label{eq:ws1}
\end{eqnarray}
which were already obtained in Ref.~\cite{Penalva:2019rgt},
 but for real Wilson coefficients, and
%
\begin{eqnarray}
\widetilde W_{I3}&=&C_T^* C_S \, F_S\left[
T_1\left(1+\omega\right)+T_3-\frac{M_{\Lambda_b}}{M_{\Lambda_c}} \left(T_2+T_4\right) \right] -
C_T^* C_P \frac{M_{\Lambda_b}}{M_{\Lambda_c}} F_P T_4, \nonumber \\ 
\widetilde W_{I4}&=&C_T^* C_V \left[F_1 T_1 \left(1+\frac{M_{\Lambda_b}}{M_{\Lambda_c}}\right)+ \frac{M_{\Lambda_b}}{M_{\Lambda_c}} F_1 \left(T_3-T_2\right)-\left(F_2 +F_3 \frac{M_{\Lambda_b}}{M_{\Lambda_c}}\right) \left(\frac{M_{\Lambda_b}}{M_{\Lambda_c}} \left(T_2+T_4\right)- T_1(1+\omega)-T_3\right)\right] \nonumber \\
&& -C_T^* C_A\frac{M_{\Lambda_b}}{M_{\Lambda_c}}\left[G_1\left(T_2+T_3\right) + 
T_4\left(G_2+ \frac{M_{\Lambda_b}}{M_{\Lambda_c}} G_3\right) \right ],\nonumber \\ 
\widetilde W_{I5}&=&-C_T^* C_V
\frac{M_{\Lambda_b}}{M_{\Lambda_c}}\left[F_1\left(T_1+T_3\right)+F_3\left(T_1\left(1+\omega\right)+T_3-
\frac{M_{\Lambda_b}}{M_{\Lambda_c}}\left(T_2+T_4\right)\right)\right]+ C_T^* C_A
\frac{M_{\Lambda_b}}{M_{\Lambda_c}} \left(G_1 T_3+ \frac{M_{\Lambda_b}}{M_{\Lambda_c}}G_3
T_4\right),
\nonumber \\ 
\widetilde W_{I6}&=&-C_T^* C_V\, F_1\left[ \left(1-\omega\right)\left(T_2+T_3\right)+ \left(1-
\frac{M_{\Lambda_b}}{M_{\Lambda_c}}\right)T_4\right]-C_T^* C_A\, G_1\left[
\frac{M_{\Lambda_b}}{M_{\Lambda_c}} \left(T_2+T_4\right)+T_4+\omega\left(T_3-T_2\right)-
\frac{M_{\Lambda_c}}{M_{\Lambda_b}}T_3\right],\nonumber \\ 
\widetilde W_{I7}&=&C_T^* C_V\, F_1 \left[T_3 \left(1-\omega \right) 
- \frac{M_{\Lambda_b}}{M_{\Lambda_c}} T_4 \right]+C_T^* C_A\, G_1
\left[ \frac{M_{\Lambda_b}}{M_{\Lambda_c}} \left(T_2+T_4\right)+\omega T_3\right],\nonumber \\ 
\widetilde W_{2}^T&=& \frac{|C_T|^2}{2}\Bigg\{\frac{M_{\Lambda_c}^2}{M_{\Lambda_b}^2}
\left[ (1+\omega)T_1^2 +2 T_1T_3\right]
-\frac{2M_{\Lambda_c}}{M_{\Lambda_b}}\left[\omega(\omega+1)T_1^2 +T_1 
\left(T_2 + 2 \omega T_3  + T_4\right) + T_2 T_3\right]\nonumber\\
&&\hspace{1.5cm}+ (1+\omega) T_1^2 - 2 \omega \left(T_2^2 + T_3^2\right)  + 2 T_1 \left(2 \omega (T_2 + T_4) + T_3\right)  
+ 2 \left(T_2 + T_3) (T_2 + T_3 + 2 T_4\right),\nonumber \\
&&\hspace{1.5cm}-\frac{2M_{\Lambda_b}}{M_{\Lambda_c}}\left[ T_1 (T_2+T_4)+T_2 (T_3+2 T_4)+2 T_4 (T_3+T_4)\right]
\Bigg\} 
\nonumber \\
 \widetilde W_{3}^T&=&\frac{|C_T|^2}{2}\left\{(1+\omega ) T_1^2  + 2 T_1 T_3 +2
 (1-\omega)T_3^2-\frac{2M_{\Lambda_b}}{M_{\Lambda_c}}\left[ T_1 (T_2+T_4)+T_3 (T_2+2
 T_4)\right]\right\},\nonumber \\
 \widetilde W_{4}^T&=&\frac{|C_T|^2}{2}\left\{ 
 \frac{M_{\Lambda_c}}{M_{\Lambda_b}}\left[ \omega (1+ \omega) T_1^2 + 
 2 \omega T_1 T_3\right] -T_1^2 (1+\omega) -2\omega (T_2 + T_4)T_1 - 
 2 T_3 \left[T_1 +T_2 +(1-\omega) T_3+ T_4\right] \right.\nonumber\\
 &&\hspace{1.5cm}\left. + \frac{2M_{\Lambda_b}}{M_{\Lambda_c}}\left[T_4 \left(T_1+T_2+2 T_3\right)+T_2 (T_1+T_3)+T_4^2 \right]
 \right\}.
\label{eq:ws2}
\end{eqnarray}
In addition, as discussed in Appendix~\ref{sec:appH}, $ \widetilde W_5^T = 0$ and, if necessary, $\widetilde W_1^T$ can be obtained from $\widetilde W_{2,3,4}^T$ using Eq.~\eqref{eq:relacWTs}.

\bibliography{B2Dbib}

\end{document}